\documentclass[fleqn,10pt,table]{wlscirep}

\usepackage{nameref}

\usepackage[utf8]{inputenc}
\usepackage[T1]{fontenc}
\usepackage{lineno}

\usepackage{pdflscape}
\usepackage{siunitx}
\usepackage{ccfonts}
\usepackage{pdflscape}
\usepackage{multirow}
\usepackage{textcomp}

\usepackage{tabularx}
\usepackage{array}
\usepackage{booktabs}

\usepackage{pifont}
\newcommand{\cmark}{\ding{51}}%
\newcommand{\xmark}{\ding{55}}%

\usepackage{soul}
\usepackage{xargs}
\usepackage[colorinlistoftodos,prependcaption,textsize=tiny]{todonotes}
\newcommandx{\improvement}[2][1=]{\todo[linecolor=violet,backgroundcolor=violet!25,bordercolor=violet,#1]{#2}}
\usepackage[T1]{fontenc}
\usepackage{lmodern}
\title{Making Mathematical Research Data FAIR: \\
A Technology Overview}

\author[1,*]{Tim Conrad}
\author[1]{Eloi Ferrer}
\author[2,3]{Daniel Mietchen}
\author[1]{Larissa Pusch}
\author[2]{Johannes Stegmüller}
\author[2]{Moritz Schubotz}
\affil[1]{Zuse Institute Berlin, Berlin, Germany}
\affil[2]{FIZ Karlsruhe, Berlin, Germany}
\affil[3]{Ronin Institute for Independent Scholarship, Montclair, United States}
\affil[*]{Corresponding author: Tim Conrad (conrad@zib.de)}

\begin{abstract}
The sharing and citation of research data is becoming increasingly recognized as an essential building block in scientific research across various fields and disciplines. Sharing research data allows other researchers to reproduce results, replicate findings, and build on them. Ultimately, this will foster faster cycles in knowledge generation. Some disciplines, such as astronomy or bioinformatics, already have a long history of sharing data; many others do not. The current landscape of so-called research data repositories is diverse. This review aims to perform a technology review on existing data repositories/portals with a focus on mathematical research data.
\end{abstract}
\begin{document}

\flushbottom
\maketitle

\thispagestyle{empty}

%%%%%%%%%%%%%%%%%%%%%%%%%%%%%%%%%%%%%%%%%%%%%%%%%%%%%%%%%%%%%%%%%%%%%%%%%%%%%%%%%%%%%%%%%%%%%%%%%
\section*{Introduction}
%%%%%%%%%%%%%%%%%%%%%%%%%%%%%%%%%%%%%%%%%%%%%%%%%%%%%%%%%%%%%%%%%%%%%%%%%%%%%%%%%%%%%%%%%%%%%%%%%

% Bercic et al. \cite{Bercic2020} recently compiled an overview (see table \ref{fig:math_data_overview}).
The importance of sharing and citing research data is steadily gaining recognition as a foundational element in scientific research across different fields and subjects. Sharing research data allows other researchers to reproduce results and replicate findings\cite{Piwowar2007,Tenopir2015}. Ultimately, this promotes the generation of knowledge at a faster pace. Some disciplines already have a long history of sharing data and are benefiting from it\cite{Lebo2018}, but many others do not.

Although the term \textit{data sharing} is not used unambiguously in the literature\cite{Thoegersen2022}, technically, data sharing is mainly organized through Research Data Repositories (RDR). The current RDR landscape is diverse. However, it can be roughly classified into four categories: institutional, disciplinary, multidisciplinary, and project-specific\cite{pampel2013making}. 

In the field of mathematics, significant progress has already been made in terms of research data sharing. Typical data types include theorem libraries or number sequences (see Table~\ref{tab:research_data_type} for more examples). As a highly structured and rigorous field, mathematics fits well with the development of shared data resources. Particularly, theorems and proofs can be conveniently disseminated and checked using available checking engines~\cite{DBLP:journals/pami/GreinerPetterSBSAG23, DBLP:conf/mkm/CohlGS18}. Overall, the available sites and repositories provide mathematicians with a large assortment of mathematical objects that can be utilized to solve new problems, establish new theories, and increase knowledge - not only in mathematics.

However, despite the advancements made in the mathematics community, data sharing is still a subject that requires continuous attention and improvement. Researchers, institutions, and funding agencies must prioritize the development of rules and infrastructure that facilitate the sharing and citation of research data to increase the prevalence of good data-sharing practices in mathematics - and other fields. By doing so, we may develop an environment for research that is more open and collaborative, thereby accelerating the rate of scientific discovery.

Also in other fields, data sharing enables more transparent and repeatable scientific research, making it an increasingly important topic in recent years. By releasing data openly, researchers can increase the likelihood that their findings can be repeated and validated by others, so bolstering the credibility of the results. However, issues persist in ensuring that data is shared in an efficient and accountable manner. To ensure its integrity and usability, the shared data must be carefully curated and documented. Caution needs also to be taken such that privacy and confidentiality issues are addressed to prevent misuse or misinterpretations of the data. By establishing best practices for data sharing and citation and facilitating the creation of standardized metadata and data management standards, these challenges can be overcome.

In conclusion, the sharing and citation of research data is a crucial part of scientific research that has the ability to boost collaboration and speed the production of new knowledge. The field of mathematics has made substantial progress in this area, but additional efforts are required to guarantee that data sharing becomes a generally accepted and well-supported practice in all fields.

%%%%%%%%%%%%%%%%%%%%%%%%%%%%%%%%%%%%%%%%%%%%%%%%%%%%%%%%%%%%%%%%%%%%%%%%%%%%%%%%%%%%%%%%%%%%%%%%%
\subsection*{Objectives \& Outline}

Through this technology review, we sought to answer the following research questions, with a particular focus on mathematical research data. 

\begin{enumerate}
    \item What is the current status of open data platforms in academia?
    \item What are the main requirements for an open data platform? 
    \item What are the biggest challenges and obstacles that are preventing the successful implementation of widely used open data platforms?
\end{enumerate}

\noindent We have structured the paper as follows: We first give the necessary background and emphasize the significance of open data platforms in mathematical research and their role in promoting Open Science in the following section. In the following methods section, we describe our methodology for compiling and evaluating a comprehensive list of mathematics-related open data platforms. The open data platforms that made it to the final list are described in the results section, in which we provide a comprehensive analysis, assessing their features and conformance to the FAIR principles. The discussion section provides an analysis of the results, highlighting the challenges and limitations of the existing open data platforms.  In the conclusion part of this paper, we provide guidelines for authors submitting to research data repositories.

%%%%%%%%%%%%%%%%%%%%%%%%%%%%%%%%%%%%%%%%%%%%%%%%%%%%%%%%%%%%%%%%%%%%%%%%%%%%%%%%%%%%%%%%%%%%%%%%%
\section*{Open Science in Mathematics: The Role of Open Data and Data Platforms}
%%%%%%%%%%%%%%%%%%%%%%%%%%%%%%%%%%%%%%%%%%%%%%%%%%%%%%%%%%%%%%%%%%%%%%%%%%%%%%%%%%%%%%%%%%%%%%%%%

Not only has the digital revolution altered how we conduct research, but also how we share it \cite{boulton2012}. In this regard, the Open Science movement, which advocates for the accessibility and reuse of scientific research, has been a game-changer. It has played a crucial role in promoting openness, effectiveness, and collaboration within the scientific community~\cite{mckiernan2016open}. The main goal is to increase the use of research results by society, industry, and science itself, thereby making the scientific community more transparent and efficient. Open access to scientific publications, open-source software, open data, and free educational materials are essential components of Open Science.

The concept of open data, which emphasizes making research data publicly accessible, is fundamental to Open Science. This practice facilitates not only the replication and validation of research findings, but also the exploration of new research questions and hypotheses.

In the context of mathematics, open data takes on a significance of its own. A variety of data types, including symbolic formulae, numerical arrays, and observational data, characterize mathematical research~\cite{DBLP:phd/dnb/Schubotz17}. Understanding these data types is indispensable for the efficient analysis and communication of mathematical research. This section delves deeper into the implications of open data in mathematical research and investigates how open data platforms can be utilized to make such data accessible to a larger audience.

%%%%%%%%%%%%%%%%%%%%%%%%%%%%%%%%%%%%%%%%%%%%%%%%%%%%%%%%%%%%%%%%%%%%%%%%%%%%%%%%%%%%%%%%%%%%%%%%%
\subsection*{Open Data in Mathematics}

As a central component of Open Science, open data refers to the practice of making research data publicly accessible under open licenses. This practice facilitates the replication and validation of findings by allowing other researchers to verify and expand upon previous research. In addition, open data enables the investigation of new research questions and hypotheses, as well as the combination of data from multiple sources to uncover novel insights and patterns. As a result, open data is becoming the norm in an increasing number of scientific disciplines.

The field of mathematical research provides an intriguing example in this regard. Numerous data types, including symbolic formulae and theorems, numerical arrays, and observational information, characterize mathematical research (see Table \ref{tab:math_research_data} for an overview). Understanding these various data types is essential for analyzing and communicating mathematical research effectively.

With a focus on mathematical research data, we will investigate open data platforms and how they can be utilized to make such data accessible to the general public. By understanding the advantages and disadvantages of open data, researchers can make well-informed decisions regarding how to share their research and contribute to the expanding Open Science movement.

\begin{table}[!ht]
\centering

\newcolumntype{R}{>{\raggedright\arraybackslash}p{8cm}}
\begin{tabularx}{16cm}{p{3cm} p{3.5cm} R }
\toprule
Category & Types & Description \\
\midrule

Symbolic data % play a crucial role in the development of new mathematical theories and the verification of existing ones. Symbolic data can also be used to generate new conjectures and insights in a wide range of mathematical fields.
% E.g. contained in Computer Algebra Systems such as SageMath
 & \raggedright Formulae, Theorems, Proofs, Functions 
 & Mathematical expressions, theorems, proofs, and functions represented using symbolic notation. \\ 
\hline
Numeric data % often used in optimization problems and in the development of algorithms and models.
 & \raggedright (Integer) number sequences, Matrices, Tensors, Finite lattices
 & Numerical values, matrices, lattices used in mathematical modeling and analysis. \\
\hline
Geometric data % used in fields such as topology, differential geometry, and algebraic geometry.
 & \raggedright Curves, Surfaces, High-dimensional objects, Polytopes
 & Objects and structures used in geometry and related fields, such as curves, surfaces, and high-dimensional objects, e.g. manifolds. \\
\hline 
Models 
 & \raggedright Math models, BioModels
 & Abstractions of real-world phenomena used to make predictions and test hypotheses. \\
\hline
Observational data 
 & \raggedright Simulations, Experiments, Observations 
 & Empirical data collected through experiments, simulations, and observations of natural phenomena. \\
\hline
Text data 
 & \raggedright arXiv.org, EuDML, Encyclopedia of Math
 & Written and digital sources of mathematical research, including papers, books, and online resources. \\
\bottomrule
\end{tabularx}
\caption{\label{tab:research_data_type}Types of mathematical research data}
\label{tab:math_research_data}
\end{table}

%%%%%%%%%%%%%%%%%%%%%%%%%%%%%%%%%%%%%%%%%%%%%%%%%%%%%%%%%%%%%%%%%%%%%%%%%%%%%%%%%%%%%%%%%%%%%%%%%
\subsection*{Open Data Platforms}
\label{sec:requirements-open-data-platform}

In recent years, data sharing has become an essential component of scientific research, as it enables researchers to increase the impact of their work and promote transparency and collaboration. Open data platforms are digital environments in which scientists can store, exchange, and access datasets. Usually, these platforms include data management tools, metadata standardization, and version control. Zenodo and Figshare are prominent open data platforms, each with its own set of features and user communities. 

Nonetheless, the process of data sharing poses various challenges, such as ensuring accessibility and the need for effective metadata management and standardization. In this regard, an open data platform serves as a centralized repository for storing and sharing (research) data, thereby offering a solution to these challenges. Furthermore, an effective open data platform offers a range of features that facilitate data sharing and reuse, including easy accessibility, enhanced discoverability, simplified data submission mechanisms, functionalities for metadata management, and conformance with FAIR principles. The fundamental key features that an open platform should provide are:

\begin{enumerate}
    \item Free use: The open data platform should be free to use for researchers, allowing them to share and access data without financial barriers.
    
    \item Accessibility: Researchers should be able to access the data from any location and computational environment, via a standard web browser. Easy accessibility promotes transparency, facilitates reproducibility, and helps to avoid duplication of efforts.

    \item Data submission mechanisms: Researchers should be able to submit data to the repository, making it available for future use and replication. This feature promotes the sharing of data and transparency of research results.
\end{enumerate}

\noindent Moreover, an effective open platform should aim to fulfill additional features such as:

\begin{enumerate}
    \item FAIR principles compliance: The platform should adhere to the FAIR principles, which emphasize the importance of data being Findable, Accessible, Interoperable, and Reusable. These principles place particular importance on the ability to process data by machines and are listed in the Section on \hyperref[sec:fair-principles]{FAIR principles}.

    \item Data quality: The platform should ensure that the data is of high quality, reliable, and accurate. This feature is critical to ensure that research data is useful and impactful for future research.

    \item Metadata management: The platform should provide tools for metadata management, including descriptions of the data, authors, and institutions. Metadata helps researchers to discover, access, and understand the data.

    \item (Meta)data format and standardization: The data platform should support various data formats and adhere to standardization guidelines to ensure the data is easily findable, accessible, interoperable, and reusable.

    \item Security and privacy: The platform should ensure the security and privacy of the data, protecting sensitive information from unauthorized access and misuse.
    
    \item User-friendly interface: The platform should be easy to use and accessible, with low entry barriers, enabling scientists from diverse backgrounds to participate in the Open Science movement. This feature includes clear and concise documentation, as well as intuitive interfaces to upload and retrieve data.
\end{enumerate}

In the following, we will describe and review existing research data platforms along those features (see \hyperref[sec:existing-platforms]{Results Section}). Keep in mind that in this review, however, we will restrict our focus on platforms that contain significant mathematical research data. Before we dive into these details, we first introduce how the data was collected.

%%%%%%%%%%%%%%%%%%%%%%%%%%%%%%%%%%%%%%%%%%%%%%%%%%%%%%%%%%%%%%%%%%%%%%%%%%%%%%%%%%%%%%%%%%%%%%%%%
\section*{Methods}
\label{sec:methods}
%%%%%%%%%%%%%%%%%%%%%%%%%%%%%%%%%%%%%%%%%%%%%%%%%%%%%%%%%%%%%%%%%%%%%%%%%%%%%%%%%%%%%%%%%%%%%%%%%

To assess the current status of open data platforms in the field of mathematics, the first challenging step consisted in obtaining a comprehensive list of all relevant platforms in the field, operational at the time of writing this review. A systematic search based only on a review of publications is not a feasible approach in this case, since most open platforms exist without being explicitly documented in the technical literature. Instead, most of the current open platforms are only findable through search engines. Thus, our approach to get an overview of the current ecosystem had to combine a literature review with the direct results obtained in a search engine. 

We started our search by reviewing published articles found in Google Scholar obtained by combining the terms “mathematics”, “research data”, “scientific data”, “research metadata”, “scientific metadata”, “portal”, “repository”, “infrastructure”, “platform”, “metadata management” and “FAIR”. By examining these publications as well as the URLs mentioned in them we obtained a first tentative list of open platforms, not only in the field of mathematics. We also identified further publicly-accessible portals directly through search engines using the same keywords and through searches in aggregators of data repositories that included \href{https://fairsharing.org/}{FAIRsharing}\cite{sansone2019fairsharing}, \href{https://mathdb.mathhub.info/}{MathHub}~\cite{iancu2014system}, \href{https://v2.sherpa.ac.uk/opendoar/}{OpenDOAR} and \href{https://www.re3data.org/}{re3data}\cite{Re3data}. The list was finally completed based on the authors’ knowledge. The initial search was performed during the period June 2022 -- October 2022. A second round of searches took place in the period March 2023 -- June 2023. It is also important to note that as a restriction on all inquiries, the availability of English-language content was mandated.

From the initial list of platforms, we excluded those that did not meet the essential requirements outlined in section on \hyperref[sec:requirements-open-data-platform]{Open Data Platforms}. Specifically, we discarded platforms that were not free to use, not publicly accessible, or did not offer the option of data submission. Not offering a direct and open mechanism for data submission is often a requirement for research data repositories that have strict requirements on data curation. For this review, this forced us to exclude known resources in the mathematical community such as the \href{https://faculty.evansville.edu/ck6/encyclopedia/etc.html}{Encyclopedia of Triangle Centers}\cite{narboux2016towards}, the \href{https://www.graphclasses.org/}{ISGCI} (Information System on Graph Classes and their Inclusions)\cite{de2016information}, the \href{http://www.grdb.co.uk/}{Graded Ring Database}\cite{brown2015graded}, the \href{https://www.lmfdb.org/}{L-functions and modular forms database}\cite{cremona2016functions} as well as a long list of publicly-accessible institutional repositories that only accept submissions by its members such as ATLAS of Finite Group Representations \url{https://brauer.maths.qmul.ac.uk/Atlas/v3/} or MIZAR \url{http://mizar.org/library/}.

Platforms that just aggregate metadata and thus do not offer a mechanism to directly incorporate data and metadata submitted by users were also excluded from the review. These include aggregators such as \href{https://re3data.org}{re3data}, \href{https://datacite.org/}{DataCite}\cite{neumann2014datacite} and \href{https://dimensions.ai}{Dimensions.ai}\cite{bode2018guide}. We also excluded open platforms that did not include a significant amount of mathematical research data at the time of this review and were thus not relevant for our purposes. Among these platforms were \href{https://b2share.eudat.eu/}{B2share}\cite{ardestani2015b2share}, \href{https://datadryad.org/}{Dryad}\cite{white2008dryad}, \href{https://fairdomhub.org/}{Fairdomhub}\cite{wolstencroft2017fairdomhub}, \href{https://data.mendeley.com/}{Mendeley Data}\cite{bhoi2018mendeley} and \href{https://vivli.org/}{Vivli}\cite{li2019moving}.

Finally, within the mathematical ecosystem, there exists a distinct category of portals that mainly contain written articles that describe concepts within specific mathematical disciplines. These platforms, often built on the MediaWiki framework, encourage user contributions through a wiki-based approach. However, they are often predominantly populated only by a small group of active contributors within the field. While these platforms are free to use, accessible, and allow user contributions, they often display limited adherence to the FAIR principles. Notably, they frequently lack unique persistent identifiers for published articles, do not provide access to comprehensive metadata via an API, and often lack explicit license information. Due to these limitations, we chose not to include them in this review, as their inclusion would have significantly expanded the list with numerous items that only marginally comply with the FAIR principles. Nevertheless, given their significance in the field of open-access mathematical research data, we have included a non-exhaustive list of such platforms in~\hyperref[sec:mediawiki-math]{Appendix B}.

The final list of platforms is included in Table~\ref{tab:overview}. Each platform has been evaluated based on Austian et al.~\cite{Austin2016Research} using publicly available information, in the following categories (see Table~\ref{tab:evaluation_criteria}): Infrastructure, Preservation, Security / Privacy, Archiving, Submission, Access / Sharing, Policy, and whether they are compliant with the FAIR principles~\cite{wilkinson2016fair}. This allowed us to identify presently implemented standards and features related to the management and sharing of research data, with a focus on mathematics. An analysis of the FAIR compliance for each platform is included in Table~\ref{tab:fair_compliance}. The following section provides also a brief description of each included platform, based on the evaluation criteria. Together, this information serves as the basis for the discussion on the presented research questions.

\newpage
\begin{landscape}

\begin{table}[!bh]
\centering
\begin{tabularx}{23cm}{p{2.5cm} p{4cm} p{15cm}}
\toprule
Category & Sub-category & Options  \\
 \midrule
Infrastructure
   & Platform  & Dataverse | Figshare | MediaWiki | Proprietary | Open-Source (other) \\
   & Cost      & Free | Free to access, but contribution needed for deposit \\
   & Size      & Size of repository (number of datasets) \\
\hline
Preservation
   & Redundancy & None | Multiple redundant copies | Geographically distributed redundant copies \\
   & Persistent identifiers & No ID | DOI | other persistent ID | non-persistent ID \\
   & Persistent data deposit & None | Long term data preservation \\
\hline
Security / 
   & Security & None | Authentication mechanism \\
Privacy
   & Privacy & None | Distinction between public and private data \\
\hline
Archiving
   & Author identifier & None | zbMATH Open Author ID | ORCID ID | SCOPUS ID | Other \\
   & Publication identifier & None | zbMATH Open Document ID | Reference to paper through DOI | Other reference \\
   & Time stamping & None | Timestamp upon upload | Timestamp for every version \\
\hline
Submission
   & Data types & No restriction | Restrictions to specific types \\
   & Data size & No restriction | Restricted to maximal size \\
   & Metadata & No metadata necessary | Controlled Language | Readme file \\
   & Review/Data Quality & None | Submissions are reviewed and approved for metadata and compliance \\
\hline
Access / Sharing
   & Online access & Data available for free and open download | User registration needed \\
   & API & None | API for search available | API for search and submission available \\
   & License & CC0 |  Creative Commons License | Other license (open) | Other license (restrictive) \\
\hline
Policy
   & Mandate & No | Yes (Info about: Under what authority does the repository operate (e.g. government)?) \\
   & Data Ownership & No | Yes (Info about: Who owns the ingested data?) \\
   & Data Licensing & No | Yes (Info about: How are the data licensed?) \\
   & Preservation & No | Yes (Info about: What is the practice for long-term preservation?) \\
   & Succession plan & No | Yes (Info about: What actions will be taken if the repository is closed?) \\
\hline
FAIR Principles
   & Findability & No | Yes (Means: The data can be discovered by both humans and machines, for instance by exposing metadata and keywords to search engines) \\
   & Accessibility & No | Yes (Means: The (meta-)data is archived in long-term storage and can be made available using standard technical procedures) \\
   & Interoperability & No | Yes (Means: The data can be exchanged and used across different applications and systems) \\
   & Reusability & No | Yes (Means: The data is well documented and licensing information is provided \\
\bottomrule
\end{tabularx}
\caption{Evaluation criteria, based on Austian et al.~\cite{Austin2016Research} and Wilkinson et al.~\cite{wilkinson2016fair}}
\label{tab:evaluation_criteria}
\caption*{}
\end{table}

\end{landscape}
\newpage

%%%%%%%%%%%%%%%%%%%%%%%%%%%%%%%%%%%%%%%%%%%%%%%%%%%%%%%%%%%%%%%%%%%%%%%%%%%%%%%%%%%%%%%%%%%%%%%%%
\section*{Results}
% Check out Moritz lit.rev.: https://www.frontiersin.org/articles/10.3389/fbloc.2019.00016/full
%%%%%%%%%%%%%%%%%%%%%%%%%%%%%%%%%%%%%%%%%%%%%%%%%%%%%%%%%%%%%%%%%%%%%%%%%%%%%%%%%%%%%%%%%%%%%%%%%

This section examines the current state of Open Science and Open Data platforms by reviewing the available literature and the implementation details of existing platforms. The evaluation is based on the criteria presented in Table~\ref{tab:evaluation_criteria} and on the adherence to the FAIR principles. Apart from a short description of each platform, summarized in Table~\ref{tab:overview}, a comparison in terms of the most relevant features for data sharing has been included in Table~\ref{tab:features}.

%%%%%%%%%%%%%%%%%%%%%%%%%%%%%%%%%%%%%%%%%%%%%%%%%%%%%%%%%%%%%%%%%%%%%%%%%%%%%%%%%%%%%%%%%%%%%%%%%
%\subsection*{Literature}
%\label{sec:existing_literature}
%\begin{enumerate}
%    \item Comparative review of integrated data repositories in Health Care Institutions: \href{https://formative.jmir.org/2020/8/e17687}{link}
%    \item \cite{Murphy2021tool} derives evaluation criteria from certain sets of principles (including open and FAIR ones) and applies them to a range of biomedical repositories
%    \item \cite{Austin2016Research} compares multiple research data repositories
%    \item \cite{Banzi2019Evaluation,Banzi2018Report} compares repositories for patient-level clinical data
%    \item \cite{Goh2006checklist} compares software packages for running research data repositories
%\end{enumerate}

%%%%%%%%%%%%%%%%%%%%%%%%%%%%%%%%%%%%%%%%%%%%%%%%%%%%%%%%%%%%%%%%%%%%%%%%%%%%%%%%%%%%%%%%%%%%%%%%%
\subsection*{Open Data Platforms}
\label{sec:existing-platforms}

%\href{https://docs.google.com/spreadsheets/d/1BCDjviYR2dh1dXuHXC_e6uZffc7m38qTHGPHPukIUfs/edit#gid=0}{spreadsheet}

For each platform listed, a brief description of its key features is provided, including information related to its creation, technological framework, objectives, and mathematical focus. If available, we include scientific papers and white-papers in which the systems are described. If no such paper is available, we refer directly to the website.

\paragraph{4TU Research Data}
\href{https://data.4tu.nl/}{4TU.ResearchData} is an online data repository for science, engineering and design, managed by the 4TU.ResearchData Consortium. It aims to facilitate the sharing of research datasets and guarantees their long-term access by adhering to FAIR principles. The repository has been online since 2010~\cite{cruz2018adding}, it is based on Figshare technology, and it is hosted and managed by the TU Delft Library. As of June 2023, it hosts slightly more than 8,000 items, which include 7,850 datasets and 174 software items. The vast majority of the items belong to the field of atmospheric sciences and climate studies. About 100 items are assigned to mathematical categories, including computation theory, numerical and computational mathematics, mathematical physics, applied and pure mathematics. Every uploaded dataset receives a DOI and can be assigned a license, the most popular being CC0 and CC BY 4.0. One of the most distinct attributes of this repository is its advanced functionality for software preservation, including integration with GitHub and GitLab, dedicated licenses for software and a repository sandbox for testing. Currently, the FAIR principles regarding findability, accessibility and interoperability are fully fulfilled while those concerning reusability are only partially fulfilled. 

\paragraph{Archive of Formal Proofs}
\href{https://www.isa-afp.org/}{The Archive of Formal Proofs} \cite{Blanchette2015Mining,MacKenzie2022Re} is a collection of 700 proofs from various areas of mathematics, including number theory, algebra, analysis, and geometry, among others. All included items have passed both classical peer review and have been verified by the theorem prover Isabelle\cite{Paulson1989}. The repository additionally contains proof libraries and examples for Isabelle system. The site was launched in 2004 and is maintained by the Isabelle user community. The content is organized in the style of a journal, with each article being a set of Isabelle theories and proofs accompanied by definitions, theorems and corollaries which are written in the dedicated input language Isar, and as such executable using Isabelle. Each entry is citable via a locally unique identifier string. The proofs are available under BSD and LGPL software licenses. Overall, the Archive of Formal Proofs fulfills just over half of the FAIR Principles.

\paragraph{arXiv}
\href{https://arxiv.org/}{arXiv} is an open access repository for scholarly preprints and postprints in eight subject areas, including mathematics, physics and computer science. It was founded in 1991, and it is currently maintained by Cornell University. Articles can be submitted to arXiv at no cost and are subject to a moderation process, but not peer-reviewed. As of June 2023, it is hosting well over 2 million scholarly articles organized into 32 distinct categories, with over 500,000 of them being within the field of mathematics. The platform has a strong commitment to open access, ensuring that all of its content is freely available to the public. It fully complies with the FAIR principles for findability, accessibility and reliability while only partially fulfilling the principles for interoperability. Authors can choose from several license types under which an item is made available, including various CC-BY variants, CC0 and an arXiv specific license.

%%%%%%%%%%%%%%%%%%%%%%%%%%%%%%%%%%%%%%%%%%%%%%%%%%%%%%%%%%%%%%%%%%%%%%%%%%%%%%%%%
% Overview Table
%%%%%%%%%%%%%%%%%%%%%%%%%%%%%%%%%%%%%%%%%%%%%%%%%%%%%%%%%%%%%%%%%%%%%%%%%%%%%%%%%

\newpage
\begin{landscape}

\begin{table}[ht]
\centering
\begin{tabularx}{23cm}{p{4.5cm} p{4.5cm} p{2.6cm} p{3.55cm} p{1.4cm} p{1.6cm}}

\toprule
Portal Name & Math focus & Citability &  License & \# Items & URL \\
\midrule

4TU Research Data 	          & Multidisciplinary             & DOI              & CC0 (default), CC             & $> 8300$      & \href{https://data.4tu.nl/}{data.4tu.nl} \\ \hline
Archive of Formal Proofs      & Proofs                        & URL              & BSD, LGPL                     & $\sim 700$    & \href{https://www.isa-afp.org/}{isa\mbox{-}afp.org} \\ \hline
arXiv 	                      & Multidisciplinary             & DOI, arXiv ID    & CC, arXiv perpetual           & $> 2,2$M      & \href{https://arxiv.org}{arxiv.org} \\ \hline
Biomodels                     & Mathematical models           & Model ID         & CC0                           & $> 2500$      & \href{https://www.ebi.ac.uk/biomodels/}{ebi.ac.uk/biomodels} \\ \hline
Database of Ring Theory       & Rings                         & Internal ID      & CC BY 4.0                     & $\sim 300$    & \href{https://ringtheory.herokuapp.com/}{ringtheory.herokuapp.com} \\ \hline
Encyclopedia of Graphs        & Graphs                        & Graph ID         & CC BY-NC-SA 3.0               & $> 11$M       & \href{http://atlas.gregas.eu/ }{atlas.gregas.eu} \\ \hline
Figshare                      & Multidisciplinary             & DOI              & CC, MIT, GPL, Apache          & $> 7,3$M      & \href{https://figshare.com}{figshare.com} \\ \hline
FindStat                      & Combinatorial statistics      & Internal ID      & CC BY 4.0                     & $\sim 2000$   & \href{https://www.findstat.org/}{findstat.org} \\ \hline
HAL                           & Multidisciplinary             & DOI, idHAL       & CC, copyright                 & $> 3$M        & \href{https://hal.archives-ouvertes.fr/}{hal.archives\mbox{-}ouvertes.fr} \\  \hline
Harvard Dataverse	          & Multidisciplinary             & DOI              & CC0 (default), custom         & $> 156000$    & \href{http://dataverse.harvard.edu}{dataverse.harvard.edu}   \\ \hline
MathRepo                      & Supporting material           & URL              & diverse, undefined            & $\sim 70$     & \href{https://mathrepo.mis.mpg.de/}{mathrepo.mis.mpg.de} \\ \hline
Network Repository            & Network datasets              & URL              & CC BY-SA                      & $> 6600$      & \href{https://networkrepository.com/}{networkrepository.com} \\ \hline
OEIS                          & Integer sequences             & Internal ID      & CC BY-NC 4.0                  & $> 350000$    & \href{https://oeis.org/}{oeis.org} \\ \hline
Open Science Framework        & Multidisciplinary             & DOI, Internal ID & diverse                       & $> 7$M        & \href{https://osf.io}{osf.io} \\ \hline
Open Science Library          & Multidisciplinary             & DOI              & diverse                       & $> 100$       & \href{https://codeocean.com/explore?query=Mathematics}{codeocean.com}  \\ \hline
Papers with Code              & Multidisciplinary             & URL              & CC BY-SA                      & $> 4000$      & \href{https://math.paperswithcode.com/}{math.paperswithcode.com}  \\ \hline
$\pi$-Base                    & Topological counterexamples   & Internal ID      & CC BY 4.0                     & $\sim 500$    & \href{https://topology.pi-base.org/}{topology.pi\mbox{-}base.org} \\ \hline
polyDB                        & Discrete geometric objects    & Internal ID      & undefined                     & $\sim 500$M   & \href{https://polydb.org}{polydb.org} \\ \hline
Science Data Bank	          & Multidisciplinary             & DOI, Internal ID & diverse                       & $> 7$M        & \href{https://www.scidb.cn/en}{scidb.cn} \\ \hline
SuiteSparse Matrix Collection & Sparse matrices               & Internal ID      & CC BY 4.0                     & $\sim 2800$   &  \href{http://sparse.tamu.edu/}{sparse.tamu.edu} \\ \hline
The House of Graphs           & Graphs                        & Graph ID         & Copyright                     & $\sim 22000$  & \href{https://houseofgraphs.org}{houseofgraphs.org} \\ \hline
Wikidata                      & Multidisciplinary linked data & Wikidata ID      & CC0                           & $> 100$M      & \href{https://wikidata.org}{wikidata.org} \\ \hline
Zenodo 	                      & Multidisciplinary             & DOI, zenodo ID   & diverse                       & $> 2,8$M      & \href{https://zenodo.org}{zenodo.org} \\
\bottomrule
\end{tabularx}
\caption{\label{tab:overview}List of included portals, sorted alphabetically.}
\end{table}

\end{landscape}
\newpage

%%%%%%%%%%%%%%%%%%%%%%%%%%%%%%%%%%%%%%%%%%%%%%%%%%%%%%%%%%%%%%%%%%%%%%%%%%%%%%%%%
% Features Table
%%%%%%%%%%%%%%%%%%%%%%%%%%%%%%%%%%%%%%%%%%%%%%%%%%%%%%%%%%%%%%%%%%%%%%%%%%%%%%%%%

\newpage
\newcolumntype{x}[1]{>{\centering\arraybackslash\hspace{0pt}}p{#1}}
\begin{landscape}

\begin{table}[ht]
\centering
\begin{tabularx}{21.8cm}{p{5cm} x{2.3cm} x{2.3cm} x{2.3cm} x{2.3cm} x{2.3cm} x{2.3cm}}

\toprule
Portal Name & Persistent ID & Author ID & Publication ID & Timestamping & API & Private data \\
\midrule

4TU Research Data              & \cmark (DOI) & \cmark (ORCID) & \cmark (DOI) & \cmark \cmark & \xmark        & \cmark  \\ \hline
Archive of Formal Proofs       & \xmark       & \xmark         & \cmark (DOI) & \cmark        & \xmark        & \xmark  \\ \hline
arXiv                          & \cmark (DOI) & \cmark (ORCID) & \cmark (DOI) & \cmark \cmark & \cmark        & \xmark  \\ \hline
Biomodels                      & \cmark       & \xmark         & \cmark (DOI) & \cmark \cmark & \cmark        & \xmark  \\ \hline
Database of Ring Theory        & \cmark       & \xmark         & \xmark       & \xmark        & \xmark        & \xmark  \\ \hline
Encyclopedia of Graphs         & \cmark       & \xmark         & \xmark       & \xmark        & \xmark        & \xmark  \\ \hline
Figshare                       & \cmark (DOI) & \cmark (ORCID) & \cmark (DOI) & \cmark \cmark & \cmark \cmark & \cmark  \\ \hline
FindStat                       & \cmark       & \xmark         & \cmark       & \cmark \cmark & \cmark        & \xmark  \\ \hline
HAL                            & \cmark (DOI) & \cmark (ORCID) & \cmark (DOI) & \cmark \cmark & \cmark \cmark & \xmark  \\ \hline
Harvard Dataverse              & \cmark (DOI) & \cmark (ORCID) & \cmark (DOI) & \cmark \cmark & \cmark \cmark & \cmark  \\ \hline
MathRepo                       & \xmark       & \xmark         & \cmark (DOI) & \cmark        & \xmark        & \xmark  \\ \hline
Network Repository             & \xmark       & \xmark         & \xmark       & \xmark        & \xmark        & \xmark  \\ \hline
OEIS                           & \cmark       & \cmark         & \xmark       & \xmark        & \cmark        & \xmark  \\ \hline
Open Science Framework         & \cmark (DOI) & \cmark (ORCID) & \cmark (DOI) & \cmark \cmark & \cmark \cmark & \cmark  \\ \hline
Open Science Library           & \cmark (DOI) & \xmark         & \cmark (DOI) & \cmark        & \xmark        & \cmark  \\ \hline
Papers with Code               & \cmark       & \cmark (ORCID) & \cmark (DOI) & \xmark        & \cmark \cmark & \xmark  \\ \hline
$\pi$-Base                     & \cmark       & \xmark         & \cmark (DOI) & \xmark        & \xmark        & \xmark  \\ \hline
polyDB                         & \cmark       & \xmark         & \cmark (DOI) & \xmark        & \cmark        & \xmark  \\ \hline
Science Data Bank              & \cmark (DOI) & \cmark (ORCID) & \cmark (DOI) & \cmark \cmark & \cmark \cmark & \xmark  \\ \hline
SuiteSparse Matrix Collection  & \cmark       & \xmark         & \xmark       & \xmark        & \cmark        & \xmark  \\ \hline
The House of Graphs            & \cmark       & \xmark         & \xmark       & \xmark        & \xmark        & \xmark  \\ \hline
Wikidata                       & \cmark       & \cmark         & \cmark (DOI) & \cmark \cmark & \cmark \cmark & \xmark  \\ \hline
Zenodo                         & \cmark (DOI) & \cmark (ORCID) & \cmark (DOI) & \cmark \cmark & \cmark \cmark & \cmark  \\
\bottomrule
\end{tabularx}
\begin{minipage}{1.05\textwidth}
\vspace{1em}
\caption{\label{tab:features}
Main features of included portals.\\
Persistent ID: \xmark\, No persistent ID for objects in the platform, \cmark\, Objects are identified with a persistent ID, e.g. DOI \\
Author ID: \xmark\, No reference ID for referenced authors, \cmark\, Authors are referenced with an ID, e.g. ORCID \\
Publication ID: \xmark\, No reference ID for referenced publications, \cmark\, Publications are referenced with an ID, e.g. DOI\\
Timestamping: \xmark\, No timestamp, \cmark\, Timestamp upon upload, \cmark \cmark\, Timestamp for every version\\
API: \xmark\, No API, \cmark\, API for search, \cmark \cmark\, API for search and submission\\
Private data: \xmark\, No distinction between public and private data, \cmark\, Creation of private repositories or datasets is possible
}
\end{minipage}

\end{table}

\end{landscape}
\newpage

%%%%%%%%%%%%%%%%%%%%%%%%%%%%%%%%%%%%%%%%%%%%%%%%%%%%%%%%%%%%%%%%%%%%%%%%%%%%%%%%%
% FAIR Table
%%%%%%%%%%%%%%%%%%%%%%%%%%%%%%%%%%%%%%%%%%%%%%%%%%%%%%%%%%%%%%%%%%%%%%%%%%%%%%%%%

\newpage
\newcolumntype{x}[1]{>{\centering\arraybackslash\hspace{0pt}}p{#1}}
\begin{landscape}

\begin{table}[ht]
\centering
\begin{tabularx}{23cm}{p{5cm} x{0.7cm} x{0.7cm} x{0.7cm} x{0.7cm} x{0.7cm} x{0.8cm} x{0.8cm} x{0.7cm} x{0.7cm} x{0.7cm} x{0.7cm} x{0.7cm} x{0.8cm} x{0.8cm} x{0.8cm}}

\toprule
\multicolumn{1}{l|}{} & \multicolumn{4}{c|}{Findable} & \multicolumn{4}{c|}{Accessible} & \multicolumn{3}{c|}{Interoperable} & \multicolumn{4}{c}{Reusable} \\ \midrule
\multicolumn{1}{l|}{Portal Name}  & F1   & F2   & F3   & \multicolumn{1}{c|}{F4}  & A1   & A1.1   & A1.2  & \multicolumn{1}{c|}{A2}  & I1     & I2     & \multicolumn{1}{c|}{I3}    & R1   & R1.1  & R1.2  & R1.3  \\ \midrule
4TU Research Data              & \cmark & \cmark & \cmark & \cmark      & \cmark & \cmark & \cmark & \cmark      & \cmark & \cmark & \cmark      & \cmark & \cmark & \xmark & \xmark    \\ \hline
Archive of Formal Proofs       & \xmark & \cmark & \xmark & \cmark      & \cmark & \cmark & \cmark & \xmark      & \cmark & \xmark & \xmark      & \xmark & \cmark & \cmark & \cmark    \\ \hline
arXiv                          & \cmark & \cmark & \cmark & \cmark      & \cmark & \cmark & \cmark & \cmark      & \cmark & \cmark & \xmark      & \cmark & \cmark & \xmark & \cmark    \\ \hline
Biomodels                      & \cmark & \cmark & \cmark & \cmark      & \cmark & \cmark & \cmark & \cmark      & \cmark & \cmark & \cmark      & \cmark & \cmark & \cmark & \cmark    \\ \hline
Database of Ring Theory        & \xmark & \cmark & \xmark & \cmark      & \cmark & \cmark & \cmark & \xmark      & \xmark & \xmark & \xmark      & \xmark & \xmark & \xmark & \cmark    \\ \hline
Encyclopedia of Graphs         & \cmark & \cmark & \xmark & \cmark      & \cmark & \cmark & \cmark & \xmark      & \xmark & \xmark & \xmark      & \xmark & \cmark & \xmark & \cmark    \\ \hline
Figshare                       & \cmark & \cmark & \cmark & \cmark      & \cmark & \cmark & \cmark & \cmark      & \cmark & \cmark & \cmark      & \cmark & \cmark & \xmark & \xmark    \\ \hline
FindStat                       & \cmark & \cmark & \cmark & \cmark      & \cmark & \cmark & \cmark & \xmark      & \cmark & \cmark & \xmark      & \xmark & \cmark & \xmark & \cmark    \\ \hline
HAL                            & \cmark & \cmark & \cmark & \cmark      & \cmark & \cmark & \cmark & \cmark      & \cmark & \cmark & \cmark      & \cmark & \cmark & \xmark & \cmark    \\ \hline
Harvard Dataverse              & \cmark & \cmark & \cmark & \cmark      & \cmark & \cmark & \cmark & \cmark      & \cmark & \cmark & \cmark      & \cmark & \cmark & \xmark & \xmark    \\ \hline
MathRepo                       & \xmark & \xmark & \xmark & \cmark      & \cmark & \cmark & \cmark & \cmark      & \xmark & \xmark & \xmark      & \xmark & \xmark & \cmark & \xmark    \\ \hline
Network Repository             & \cmark & \cmark & \xmark & \cmark      & \cmark & \cmark & \cmark & \xmark      & \xmark & \xmark & \xmark      & \xmark & \cmark & \xmark & \xmark    \\ \hline
OEIS                           & \cmark & \cmark & \xmark & \cmark      & \cmark & \cmark & \cmark & \xmark      & \cmark & \xmark & \cmark      & \cmark & \xmark & \cmark & \cmark    \\ \hline
Open Science Framework         & \cmark & \cmark & \cmark & \cmark      & \cmark & \cmark & \cmark & \xmark      & \cmark & \xmark & \xmark      & \cmark & \cmark & \xmark & \xmark    \\ \hline
Open Science Library           & \cmark & \cmark & \xmark & \cmark      & \cmark & \cmark & \cmark & \cmark      & \cmark & \cmark & \cmark      & \cmark & \cmark & \xmark & \xmark    \\ \hline
Papers with Code               & \cmark & \cmark & \cmark & \cmark      & \cmark & \cmark & \cmark & \cmark      & \cmark & \cmark & \cmark      & \cmark & \cmark & \xmark & \cmark    \\ \hline
$\pi$-Base                     & \cmark & \cmark & \xmark & \cmark      & \cmark & \cmark & \cmark & \xmark      & \xmark & \xmark & \cmark      & \xmark & \xmark & \xmark & \cmark    \\ \hline
polyDB                         & \cmark & \cmark & \xmark & \cmark      & \cmark & \cmark & \cmark & \xmark      & \cmark & \cmark & \xmark      & \xmark & \xmark & \xmark & \cmark    \\ \hline
Science Data Bank              & \cmark & \cmark & \cmark & \cmark      & \cmark & \cmark & \cmark & \cmark      & \cmark & \cmark & \xmark      & \cmark & \cmark & \xmark & \xmark    \\ \hline
SuiteSparse Matrix Collection  & \cmark & \cmark & \xmark & \cmark      & \cmark & \cmark & \cmark & \xmark      & \xmark & \xmark & \xmark      & \xmark & \cmark & \xmark & \xmark    \\ \hline
The House of Graphs            & \cmark & \cmark & \xmark & \cmark      & \cmark & \cmark & \cmark & \xmark      & \xmark & \xmark & \xmark      & \xmark & \xmark & \xmark & \cmark    \\ \hline
Wikidata                       & \cmark & \cmark & \cmark & \cmark      & \cmark & \cmark & \cmark & \cmark      & \cmark & \cmark & \cmark      & \cmark & \cmark & \cmark & \cmark    \\ \hline
Zenodo                         & \cmark & \cmark & \cmark & \cmark      & \cmark & \cmark & \cmark & \cmark      & \cmark & \cmark & \cmark      & \cmark & \cmark & \xmark & \xmark    \\ \hline

\end{tabularx}
\caption{\label{tab:fair_compliance}FAIR compliance of included portals, sorted alphabetically.}
\end{table}

\end{landscape}
\newpage

\paragraph{Biomodels} 
The European Bioinformatics Institute (EBI) hosts the \href{https://www.ebi.ac.uk/biomodels/}{BioModels}~\cite{malik2020biomodels} platform, an open data resource that provides access to more than 1000 curated computational models in systems biology. These models are derived from descriptions of biological phenomena found in the scientific literature, ranging from molecular and cellular processes to more complex views of whole organisms. Each model is assigned a unique and permanent identifier that can be used to cite the models. The platform supports interoperability by permitting models to be downloaded in various formats, such as SBML (Systems Biology Markup Language) or as an ODE system in the Octave syntax. BioModels provides a comprehensive and easily accessible compilation of mathematical models that describe biological systems, thereby promoting research, replication, and collaboration. All models are shared through the CC0 license.

\paragraph{Database of Ring Theory}
The \href{https://ringtheory.herokuapp.com/}{Database of Ring Theory} is a comprehensive collection of rings and modules. It was created in 2013 by Ryan C. Schwiebert and currently holds a total of 162 rings, which can be explored through a list of 175 properties. It also stores data on 11 modules, classified according to 51 distinct properties and 63 theorems that are classified into 8 categories. Each object within the database can be accessed through its URL, which contains an internal ID. While the data within this portal is also published in a repository on GitHub~\cite{Schwiebert_RingApp_2023}, it is important to note that the license information, specifically the CC BY 4.0 license, is not explicitly mentioned on the website. However, this database encourages user participation by enabling data submissions through pull requests on the associated GitHub repository. The current implementation of the database has limited compliance with the FAIR principles.

\paragraph{Encyclopedia of Graphs}
The \href{http://atlas.gregas.eu/}{Encyclopedia of Graphs} is an online repository of graph collections established in 2012 as part of the GreGAS project, funded by the European Science Foundation. As of 2023, it holds 46 collections that encompass not only graphs but also graph-like structures such as maps, configurations, and networks. Users can filter the objects in each collection using a list of over 30 distinct properties that vary depending on the specific collection. Each graph is assigned a unique Universal Graph Identifier, enabling direct access to its properties. The graph data within the repository is provided in the canonical sparse6 format and is released under the \mbox{CC BY-NC-SA 3.0} license. The platform partially complies with the FAIR principles, but lacks interoperability due to the absence of machine-readable metadata, accessible through an API.

\paragraph{Figshare}
\href{https://figshare.com/}{Figshare} is a general purpose scientific repository operated by commercial UK-based company Digital Science \& Research Solutions Ltd. It was established in 2011 and supports researchers from all disciplines\cite{Thelwall2016figshare}. As of June 2023, it contains more than 7 million records, including around 350,000 entries for mathematics. The mathematical records consist mainly of figures, datasets, and journal articles. Figshare offers generation of DOI's for hosted content, and various licenses are available, such as the various Creative Commons licenses, GPL variants, Apache and MIT licenses. Furthermore, Figshare offers a public REST-based API, OAI-PMH endpoints, and the possibility to integrate GitHub, GitLab, Overleaf and other applications. Figshare fulfills the FAIR principles for findability, accessibility and interoperability and partially fulfills the principles for reusability. For interoperability, Figshare supports OAI-PMH which enables the inclusion of qualified references in (meta-)data. The domains which define the topics of records use controlled vocabularies. A license can be selected for a record for reusability, but the selection is not required. Figshare supports a data citation metadata schema which can be customized by the users. 

\paragraph{FindStat}
\href{https://www.findstat.org/}{FindStat} is an online database dedicated to combinatorial statistics and their relations~\cite{berg2014findstat}. Inspired by the OEIS, the project was initiated in 2011 by Chris Berg and Christian Stump at the Université du Québec. Within the database, users can explore nearly 2000 combinatorial statistics, organized into 24 distinct categories, along with 296 maps and 24 collections. The platform is continuously updated with new entries that can be submitted through an online form. Each object receives a unique identifier for easy access to its properties page, explicitly mentioned in the metadata. Released under a CC BY 4.0 license, the data can be accessed in plain text or JSON format and integrates seamlessly with SageMath. FindStat fully complies with the FAIR principle of findability and partially complies with the remaining principles.

\paragraph{HAL open science}
\href{https://hal.archives-ouvertes.fr}{HAL open science} (Hyper Article en Ligne) is an open access data repository for all academic fields. It is operated by the French data center \textit{Centre pour la communication scientifique directe} (CCSD), which is part of the \textit{French National Centre for Scientific Research} (CNRS). HAL was launched in 2001\cite{Baruch2007open} and stores around three million records. Since HAL is a major french research data infrastructure, many publications are written in French. Out of this, more than 130,000 entries are connected to mathematics as of June 2023. The majority of the mathematical entries are journal articles, conference papers or preprints. HAL stores approximately 3000 entries of non-written mathematical research data such as videos, software and photos. HAL offers generation of DOIs for hosted content, and various Creative Commons licenses are available. The HAL platform fully fulfills the FAIR principles except for some reusability sub-principles. 
%What could be improved is the lack of policy texts for HAL. It has a policy mandate, but security, data ownership and a preservation plan are not available. 

\paragraph{Harvard Dataverse}
The \href{https://dataverse.harvard.edu/}{Harvard Dataverse} is a cross-disciplinary institutional repository 
open for submissions from around the world.
% Open to the world wide research data-sharing community
% , primarily for data originating from Harvard University or collaborations with it. However, data from other sources can also be found. 
% The open-source software Dataverse platform 
It runs on the open-source software Dataverse 
% Project 
% software
which has been
% is 
in operation since 2006~\cite{magazine2011dataverse} and
is maintained by the Institute for Quantitative Social Science at Harvard University. 
The Harvard Repository is one of about 100 installations of the Dataverse software, and as of June 2023, it 
hosts about 170,000 records 
% as of June 2023 
(6,000 
% so called 
Dataverse collections and 164,000 datasets). A Dataverse collection is a customizable collection of datasets (or a virtual repository) for organizing, managing, and showcasing datasets, with features allowing custom metadata and searchable metadata facet selection. 
Overall, this repository contains about 1,7 million files. About 500 datasets and 120 Dataverse collections were tagged under "Mathematical Sciences". Multiple tags are possible for a given record, and ``Computer and Information Science'' as well as ``Social Sciences'' are most frequently associated with mathematical content. The data can be put under various licenses, including Creative Commons licenses and the Creative Commons Zero (CC0) waiver. The Harvard Dataverse meets most FAIR criteria.
%\footnote{In this context, the term ``dataverse'' has at least four distinct meanings: (i) the Dataverse software that provides repository functionality \cite{King2007}, (ii) an individual instance of that software, in this case the Harvard Dataverse, (iii) the container format dataverse that can contain individual files, one or more datasets or other dataverses, (iv) the Dataverse Project that is an organizational framework supporting and guiding the development of the software and the infrastructures and communities around it.} 

\paragraph{MathRepo}
% https://arxiv.org/abs/2202.04022
The \href{https://mathrepo.mis.mpg.de/}{MathRepo}~\cite{fevola2022mathematical}, or Mathematical Research Data Repository, is a repository for mathematical research data of and by the Max Planck Institute for Mathematics in Sciences. It has been operational since 2017 and as of 2023 contains more than 70 records. About half of the FAIR principles are not fulfilled, however, they state their plan of restructuring the website to meet the FAIR criteria and follow MaRDI recommendations. The site is built on GitLab with Read the Docs and does not provide an API, but can be queried using URL strings. There is no general license for all records; instead, contributors can choose their preferred license.

\paragraph{Network Repository}
The \href{https://networkrepository.com/}{Network Repository} is a cross-disciplinary repository for network graph data \cite{Rossi2015network}. Established in 2012, it has as of June 2023 about 6,600 networks classified in more than 30 domains, which are all available under the terms of a Creative Commons Attribution Share-Alike License (version not specified). The platform assigns a unique string identifier to each network and enables comparisons between different networks based on a given list of properties. A key feature of the site is that it offers interactive visualizations to explore the data. Users are invited to upload suitable graph data. The metadata related to each network is not available in a machine-readable format and thus the platform does not fulfill the interoperability FAIR principles. The rest of the principles are partially fulfilled.
% % Daniel sent an email to the maintainers for further information. No response received.

\paragraph{On-Line Encyclopedia of Integer Sequences}
The \href{http://oeis.org/}{On-Line Encyclopedia of Integer Sequences} (OEIS) \cite{sloane2003line} was founded in 1964 and contains integer sequences and further information about the individual items. In 1996, the corresponding website was launched. Target groups are professional as well as amateur mathematicians. One of the key features is its ability to search and compare sequences. Each sequence is identified by a serial number, which makes it unequivocally identifiable. The information provided by the encyclopedia includes the sequence itself, paper references, links to material concerning a sequence, the formula used to generate it, keywords, as well as code in several programming languages and visualizations. It presently contains more than 300,000 sequences. The data contained in the Encyclopedia are made available under the CC BY-NC 4.0 license. The OEIS fulfills most of the FAIR criteria.

\paragraph{Open Science Framework}
\href{https://osf.io/}{The Open Science Framework} (OSF)~\cite{foster2017open} is an open-source platform designed to support the entire research (-project) lifecycles. It offers functionalities to design studies, collect and analyze data as well as to publish reports and archive results. The platform is developed and maintained by the Center of Open Science (COS), a non-profit technology organization founded in 2013 that supports scientific research communities. Initially conceived for the field of psychology research, it has since its foundation become multidisciplinary. As of June 2023, it contains more than 4,500 files, 1,800 preprints and 1,300 projects in the field of mathematics. These represent only a minority among the over a million indexed preprints and over six million files hosted on the platform. For each project created using the platform, a DOI can be generated, and a license can be chosen, including Creative Commons, MIT, Apache, and GNU General Public licenses. The platform also offers file storage, version control and integration to citation management and storage tools, including Mendeley, Zotero, Figshare and GitHub. The platform adheres only partially to the FAIR principles, as it lacks persistent metadata, and does not include qualified references or detailed data provenance.

\paragraph{Open Science Library}
The \href{https://codeocean.com/explore?query=Mathematics&filter=all&refine=field}{Open Science Library} is part of the Code Ocean platform, which provides cloud-based computational environments. This allows computational researchers to share their data and the necessary code to enable others to reproduce the published analysis. It is run by the commercial company Code Ocean Inc. since 2016. The key feature of this platform is that all-needed components, i.e., data, source codes, and the runtime-environment are packaged together as a container (``compute capsule''). These containers are hosted on the platform and can be run from a web browser or locally without the need to install libraries or runtime-special environments. The platform contains more than 3000 capsules categorized into multiple research disciplines, including mathematics. A capsule is assigned a DOI, is built on FAIR principles, and allows easy migration across operating systems and platforms. Licenses vary and can be chosen by the authors (e.g. MIT, CC0, etc.).

\paragraph{Papers with Code} 
\href{https://paperswithcode.com/}{Papers with Code} is an online resource that connects scientific papers with code implementations, datasets, methods and evaluation tables. The platform offers additional valuable resources such as benchmarks that facilitate the comparison between state-of-the-art models. The entries in the platform can be explored through six specialized portals, which include the fields of machine learning, computer science, physics, astronomy, mathematics and statistics. The specific portal for mathematics contains as of June 2023 more than 4000 datasets. Open to contributions from all users, the website operates under a CC BY-SA license. Each paper is assigned a string ID based on its title. This ID can be used to access the paper metadata through a REST API. The API also allows access to metadata regarding authors, conferences, datasets, evaluations, methods, models and repositories. The platform fully adheres to the FAIR principles.

\paragraph{$\pi$-base}
\href{https://topology.pi-base.org/}{$\pi$-base} is a community database that focuses on topological counterexamples. Launched in 2014 by James Dabbs, the project has grown to include 79 spaces, each offering information through 146 properties and 344 theorems. To facilitate easy access, each object within the database is assigned a unique ID, allowing users to retrieve specific objects via their corresponding URLs. While the data is also available in a GitHub repository~\cite{pi-base-data}, it lacks direct accessibility in a machine-readable format. The GitHub repository operates under a CC BY 4.0 license, but this licensing information is not explicitly stated on the website. Additionally, the repository features a guide outlining contribution conventions, offering users guidance on how to contribute effectively.

\paragraph{polyDB}
\href{https://polydb.org/}{polyDB} is a database of discrete geometric objects which was launched in 2013 by Andreas Paffenholz and Silke Horn as an extension of the software package polymake~\cite{paffenholz2017polydb}. As of June 2023 the database contains 21 collections that are classified into four groups: Manifolds, Matroids, Polytopes and Tropical Objects. In total, these collections contain more than 500 million documents. The data for each document and collection is stored as plain JSON and can be accessed through a REST API. To this purpose, both the collections and the documents receive a unique ID that can be used to access the data. Instructions to submit new collections are provided but there is no explicit information on the license under which the data is released. The platform partially fulfills all FAIR principles except the ones related to reusability.

\paragraph{Science Data Bank}
The \href{https://www.scidb.cn/en}{Science Data Bank} (ScienceDB)\cite{Chenzan2017sciencedb} is a public multidisciplinary research data repository for eScience which was launched in 2015. ScienceDB aims to become a long-term data sharing and data publishing repository in China which covers the entire spectrum of scientific fields. As of June 2023, it has close to 6 million open datasets, with over 25,000 being related to mathematics. These mathematics-related records consist of journal publications as well as datasets, slides, code data and other multimedia content and cover a wide range of mathematical topics. Users have the option to select from a range of licenses, including CC-licenses, for licensing their published data. The uploaded data undergoes a review process by the curators and can be assigned a DOI. To facilitate access and utilization, ScienceDB provides an open REST-based API that allows users to retrieve metadata, conduct entry searches, and obtain dataset metrics. While ScienceDB aligns with the majority of the FAIR principles, it does not include qualified references in the metadata.

\paragraph{SuiteSparse Matrix Collection}
The \href{http://sparse.tamu.edu/}{SuiteSparse Matrix Collection}\cite{kolodziej2019suitesparse} is a curated set of sparse matrices that arise in real applications from a wide spectrum of domains, such as thermodynamics, material science and optimization. The target group is the numerical linear algebra community, which is provided with curated data allowing for robust and repeatable experiments or for benchmarking purposes. Matrices are identifiable by ID and related metadata, such as the matrix norm or the structural rank. The matrices can be accessed over several interfaces for Java, Matlab, Julia and Python and are made available under a CC BY 4.0 License. The SuiteSparse Matrix Collection fulfills less than half of the FAIR criteria. 

\paragraph{The House of Graphs}
\href{https://houseofgraphs.org}{The House of Graphs} (HoG) provides a searchable database of graphs and network structures. It was created in 2013 and includes a growing collection of graphs with nearly 22,000 entries~\cite{coolsaet23} that are classified based on various characteristics, such as size, degree distribution, and connectivity. Registered users can add new graphs to the database and existing graphs can be downloaded in various formats, along with their corresponding metadata. The HoG also provides tools for graph visualization, enabling researchers to gain insights into the structures of the graphs in the database. No information about the used licenses is given. The current implementation of the platform adheres to most of the FAIR principles for findability and accessibility but does not comply the principles for interoperability and reusability.

\paragraph{Wikidata}
\href{https://wikidata.org}{Wikidata} is an open, cross-disciplinary and multilingual collaborative knowledgebase
\cite{Vrandecic2014Wikidata} that has taken Wikipedia's ``anyone can edit'' approach to the Linked Open Data world. It is built on MediaWiki, with a set of extensions for handling of mathematical expressions~\cite{DBLP:conf/mkm/SchubotzW14,DBLP:conf/jcdl/SchubotzGMTG20} and structured data, collectively known as Wikibase. Launched in 2012, Wikidata currently contains about 1.5 billion statements about 100 million items and 1 million lexemes, expressed via about 10,000 properties. The over 20,000 monthly contributors have made a total of about 2 billion edits so far, mostly via semi-automated tools. The data is licensed CC0 and accessible as dumps, via APIs, via a SPARQL endpoint and via a range of tools for browsing or editing. Wikidata meets all FAIR criteria~\cite{Waagmeester2020Wikidata}. Roughly 1\mbox{\textperthousand} of the content is math-related, including math publications, mathematicians, mathematical research organizations, societies, databases, conferences, software packages, algorithms, theorems, proofs, numbers, number series and more, albeit usually with incomplete coverage\cite{DBLP:conf/semweb/ScharpfSG21,DBLP:conf/sigir/ScharpfSG18}. Wikidata's 2022 growth rate was approximately 4\% for items, 12\% for properties, 46\% for lexemes and 6\% for statements.

\paragraph{Zenodo}
\href{https://zenodo.org/}{Zenodo} \cite{https://doi.org/10.25495/7gxk-rd71} is an open science data repository maintained by CERN based on the open-source Invenio framework. It was created in 2015 to provide a solution for scientists to store, share, and publish their research data and digital artifacts, such as research papers, software, or data sets. The Zenodo system provides users with a range of services, including long-term data preservation, versioning, data citations, and DOIs. The platform has a simple and user-friendly interface, making it easy to upload and manage research data. Zenodo is integrated with a range of other platforms and services, including Github, CERN's Open Data Portal, and the European Open Science Cloud, among others. The Zenodo platform contains almost 3 million records, with the majority of them being freely accessible according to the FAIR criteria.

%%%%%%%%%%%%%%%%%%%%%%%%%%%%%%%%%%%%%%%%%%%%%%%%%%%%%%%%%%%%%%%%%%%%%%%%%%%%%%%%%%%%%%%%%%%%%%%%%
\section*{Discussion}
%%%%%%%%%%%%%%%%%%%%%%%%%%%%%%%%%%%%%%%%%%%%%%%%%%%%%%%%%%%%%%%%%%%%%%%%%%%%%%%%%%%%%%%%%%%%%%%%%

The previous section has introduced several prominent repositories for sharing mathematical research data, highlighting their distinctive characteristics. By examining the collected data for each repository, a comparison can be made in terms of their focus, size, and available features.

An important aspect to take into account is the repository's capability to assign persistent identifiers to its resources. This includes both the items stored within the repository and the references to other resources, such as authors and publications. This ability directly impacts the indexing and citation process of resources within the repository, and it is also closely tied to the FAIR principles, particularly in terms of facilitating findability.

Platforms can also be compared based on their metadata management features. In this context, it is crucial to assess whether a clear distinction is made between the data itself and the accompanying metadata. Additionally, it is important to determine whether the metadata includes relevant fields that describe the resource, such as qualified references to other resources in the form of persistent IDs. Another important feature to consider is the inclusion of timestamping data within the metadata, which can indicate when the resource was created or updated. Also relevant to the management of metadata is the platform’s provision of API endpoints for retrieving metadata, as well as search and submission capabilities. These features are directly related to the FAIR principles of interoperability.

Lastly, we have also analyzed the status of the current platforms with regard to the reusability of their data. This assessment is based on their adherence to the FAIR principles on reusability as well as on the availability of license information for the published resources.

The diverse range of features and services provided by repositories plays a crucial role in promoting open science practices and fostering collaborations among researchers. Given the importance of online data repositories in scientific research and their variations in focus, functionality and FAIR compliance, researchers should carefully select the most suitable repository based on their research needs and the nature of the data they intend to share.

Based on all the evaluated factors we present a thorough analysis of the current status of open platforms in the field of mathematics, which can support this selection process. This is accompanied by a specific analysis of the current status of FAIR compliance among the reviewed platforms. 

%%%%%%%%%%%%%%%%%%%%%%%%%%%%%%%%%%%%%%%%%%%%%%%%%%%%%%%%%%%%%%%%%%%%%%%%%%%%%%%%%%%%%%%%%%%%%%%%%
\subsection*{The Status of Open Data Platforms}

The growing number of open data platforms in science and academia indicates a significant shift toward the democratization of knowledge and the promotion of open science. These platforms provide the necessary infrastructure for storing, sharing, and reusing data, which promotes collaboration, transparency, and the advancement of research. Tables~\ref{tab:overview} and~\ref{tab:features} summarize our evaluation of the current status of these platforms and their usage, especially within mathematical research.

The focus of a platform plays a fundamental role in determining the features it offers. This is particularly important in relation to the discipline or type of data that the platform stores. As can be observed from the second column in Table~\ref{tab:overview}, platforms can be categorized into two distinct types: multidisciplinary repositories that contain a significant amount of math data and specialized platforms that exclusively store data within specific mathematical disciplines, such as abstract algebra, statistics or topology. The platform's focus determines the type of data it accepts and directly impacts decisions regarding the technology used, the need for curation and review processes, the implementation of a metadata scheme, and the policies for persistent data storage.

Within the multidisciplinary platforms, there is a subgroup that consists of general-purpose data repositories such as 4TU research data, Figshare, Harvard Dataverse, the Science Data Bank, and Zenodo. These platforms aim to facilitate the sharing of research data and are therefore targeted at a similar audience. Consequently, it is not surprising that some of these platforms are partially built on the same software, as is the case with 4TU Research data, which is built on Figshare. As a result, these platforms offer a similar range of features, including the assignment of DOIs to resources, ensuring long-term data preservation, providing options for both public and private repositories, and timestamping metadata for each uploaded version.

Another group of multidisciplinary platforms focuses on enhancing collaboration, study design, and data analysis among researchers. One example is the Open Science Framework, which serves as a general-purpose data repository but also offers functionalities to manage the entire research project lifecycle. Similarly, the Open Science Library falls into this category. Although the OSL is limited to storing data and source code packaged as computational capsules, it provides researchers with tools for study design, data collection and analysis, report dissemination, collaboration, and integration with other services.

Lastly, there is a third group of multidisciplinary platforms that exclusively house textual data in the form of scientific publications or metadata associated with them. Examples of such platforms include arXiv, HAL, and Papers with code. arXiv and HAL primarily focus on storing scientific articles, while Papers with code goes a step further by linking these articles to code repositories and available benchmarks.

On the opposite end of the spectrum, we encounter a group of platforms that specialize in storing specific types of data. Examples of these platforms include Archive of Formal Proofs, the Database of Ring Theory, the House of Graphs, and the On-Line Encyclopedia of Integer Sequences, which focus on storing proofs, ring data, graph data, and integer sequences, respectively. Typically, these platforms are built using customized code based on available frameworks. Many of these platforms also choose to open-source their source code and data, often making them available as repositories on platforms like GitHub or GitLab. This approach offers several advantages for platforms with a relatively small number of items, such as $\pi$-base, the Database of Ring Theory, and MathRepo. It provides redundancy for data storage, facilitates authentication mechanisms, and enables users to submit new content through pull requests.

Another key aspect that varies among platforms is how they manage persistent identifiers. This includes the assignment of identifiers to stored resources, as well as the inclusion of qualified references to other resources like authors and publications. While the majority of the reviewed platforms assign persistent identifiers to their resources, there are some exceptions. Platforms like the Archive of Formal Proofs, MathRepo, the Network Repository, and Papers with Code can only refer to resources using a URL, without guaranteeing persistence. In contrast, the remaining platforms all provide internal persistent IDs for their resources. Many general-purpose platforms, in addition to their internal IDs such as arXiv ID, idHAL, or Zenodo ID, also assign a DOI. The allocation of a DOI is particularly important as it simplifies citation practices and helps monitoring the use of data, as well as giving credit to data providers.

None of the reviewed platforms enforces an assignment of a persistent identifier to the resource creator. However, some platforms offer optional fields to include identifiers, such as the ORCID ID. Notably, this functionality is primarily supported in multidisciplinary platforms and not in any of the math-specific portals. Additionally, some of the general-purpose platforms generate internal author IDs on an optional basis, as seen in arXiv with the arXiv author ID. While some platforms only store author names as plain text, others like OSL or polyDB also include affiliation information. Certain platforms enable the creation of profile pages for individual authors, allowing them to voluntarily add identifiers. This feature is present in platforms based on MediaWiki, including Wikidata and OEIS.

In contrast, when it comes to referencing publications, the support for persistent identifiers is more prevalent compared to authors, as Table~\ref{tab:features} confirms. Nearly all platforms employ some form of identifier to reference other publications. The most common method is through the use of DOIs. However, there is an interesting exception in the case of FindStat, which uses MathSciNet IDs to reference the mentioned publications. Furthermore, cross-referencing based on internal identifiers from other platforms is also present. For example, both FindStat and the Encyclopedia of Graphs include references to integer sequences using the OEIS ID.

Another distinguishing factor among the platforms is their approach to data review and curation. Specialized repositories focused on specific domains prioritize data curation to ensure the publication of high-quality data. Due to their specialization, these platforms store limited types of data, making the data curation process more manageable. In contrast, general-purpose repositories that cover a wide range of disciplines and types of data often do not perform data curation or review. These platforms typically conduct some form of monitoring for uploaded datasets to ensure compliance with site policies. In some cases, they only review the uploaded metadata, as seen in the case of 4TU Research Data. Other repositories, such as Harvard Dataverse, 
% offer data curation as a paid service. 
review all deposits to ensure reusable data are included, offer free consultation services to help users set up their collections, ensure proper metadata, and offer data curation as a paid service.
In cases where only basic data control is implemented, it is also common practice to include timestamp information whenever the uploaded data is updated or modified.

The availability of API endpoints for (meta)data retrieval varies among the platforms. Notably, more than a third of them do not offer any API functionality, limiting access to their data solely through a web browser. This absence often leads to a lack of clear differentiation between data and metadata, significantly hindering data reusability. However, among the platforms that provide API endpoints for resource exploration and search, around half of them also support the submission of new resources through the API. It is worth noting that these capabilities are predominantly found in general-purpose repositories.

The licensing terms for the reviewed platforms vary considerably. Numerous platforms provide versatile licensing options, such as numerous Creative Commons (CC) licenses, BSD, LGPL, MIT, GPL, Apache, and copyright licenses. In some cases, platforms adopt a default license that can be modified by the author. For example, both 4TU Research Data and Harvard Dataverse default to CC0 for data release. Conversely, platforms like Biomodels and Wikidata strictly require the use of CC0.

Judging by the quantity of items (i.e. datasets, papers, proofs, models, sequences, etc.) available on these platforms, arXiv, Figshare, HAL, Zenodo, and the Open Science Framework are most popular within the academic community among the multidisciplinary platforms. Other platforms specializing in very specific types of data store an even larger number of items, such as Wikidata, the Encyclopedia of Graphs or PolyDB, each of which contains millions of items. 

In conclusion, the analysis of the current landscape of open data platforms shows that they have acquired significant traction in science and academia, enabling the standard, accessible sharing and reuse of data. Despite their advantages, there are still obstacles to overcome, including enhancing citability through use of persistent identifiers, standardizing licensing practices and improving interoperability through APIs. As the data landscape continues to evolve, these platforms must adapt to meet the changing needs of researchers and ensure the continued advancement of open science. In the next section, we will specifically examine the status of FAIR compliance among the reviewed platforms.

\subsection*{FAIR Compliance}
Table~\ref{tab:fair_compliance} provides an assessment of the platforms included in this review with regard to their adherence to the FAIR principles. The data gathered in this table reveals two main points. Firstly, it highlights the shortcomings regarding FAIR compliance within the current open platform ecosystem, specifically within the domain of mathematics. Secondly, it brings to attention the challenges encountered in adhering to certain FAIR principles, which are often not only technical but arise also from ambiguous interpretations of the principles and due to the absence of relevant standards within certain disciplines.

In relation to the findability principles, almost all reviewed platforms assign unique and persistent identifiers to their resources, thereby satisfying principle F1. Moreover, almost all of the reviewed platforms provide rich metadata on each object, thereby also fulfilling principle F2. It is worth noting, however, that while most platforms do establish a clear distinction between data and metadata, half of them do not explicitly reference the persistent identifier within the metadata, resulting in non-compliance with principle F3. This principle is also immediately not fulfilled by those platforms that do not assign persistent identifiers in the first place. Finally, the adherence to principle F4 concerning the indexing of the resources does not pose a technical challenge and is fulfilled by all platforms. 

Given that this review focuses on open platforms, all the platforms under examination adhere to principle A1 regarding accessibility, as well as its corresponding sub-principles. However, it is worth mentioning that only half of these platforms comply with principle A2, which guarantees the availability of metadata even if the original data resource is deleted. The non-adherence to this principle can often be attributed to the absence of a clear distinction between data and metadata, coupled in some platforms with the lack of persistent identifiers.

Adhering to the interoperability principles requires the publication of metadata in a format that can be readily interpreted by machines. In this regard, approximately two thirds of the platforms grant access to metadata via an API, but only less than half of them employ controlled vocabularies for metadata description. Furthermore, qualified references to other metadata, such as DOIs for publications or ORCID IDs for referenced authors, are included in only half of the platforms, as can also be seen in Table~\ref{tab:features}.

With regard to the principles concerning reusability, principle R1 requires enriching metadata to assess the usefulness of data within a given context. While a clear definition of rich metadata is lacking, our evaluation has determined that only half of the platforms offer some form of contextual metadata. Moreover, six of the reviewed platforms do not disclose metadata with explicit license information, thereby failing to meet the requirements for principle R1.1. Adhering to principle R1.2 presents a technical challenge due to the difficulties involved in describing data provenance and workflows by means of controlled vocabularies and machine-readable formats. This principle is easily fulfilled by platforms that exclusively store specific data types, such as descriptions of particular mathematical objects, but poses a challenge in multidisciplinary platforms capable of storing diverse data resources. For the same reason, compliance with principle R1.3, which mandates conformity to domain-relevant community standards, proves challenging for multidisciplinary platforms. The adherence to this principle relies on the existence of community standards, which may not always be present, particularly in niche disciplines, consequently impeding its compliance.

A general overview of these results show that, with the exception of principles F3 and A2, most of the platforms fulfill the FAIR principles for findability and accessibility. However, compliance rates are comparatively lower in the areas of interoperability and reusability. In particular: 

\begin{enumerate}
    \item Findability principles do not present a technical challenge and are mostly fulfilled by all platforms. The main shortcoming in the mathematical ecosystem is the absence of explicit references to persistent identifiers (F1) in the metadata, thereby preventing the fulfillment of principle F3.
    \item Accessibility principles are entirely fulfilled except for principle A2. This last principle requires a clear distinction between data and metadata, along with ensuring the availability of metadata even if the original resource is deleted.
    \item The compliance of the interoperability principles is coupled with the technical challenges of publishing rich metadata in a machine-readable format. Even among the platforms that fulfill principle I1, there is still a lack of compliance regarding the use of controlled vocabularies with persistent identifiers and the inclusion of qualified references to other metadata.
    \item The reusability principles are only partially fulfilled among the covered platforms. The compliance with these principles could be improved by adding contextual metadata, explicit license information and metadata on data provenance. The adherence to principle R1.3 requires not only a technical implementation by the platforms but also the definition and acceptance of well-defined community standards within each discipline, and in particular, in the different mathematical disciplines.    
\end{enumerate}

%%%%%%%%%%%%%%%%%%%%%%%%%%%%%%%%%%%%%%%%%%%%%%%%%%%%%%%%%%%%%%%%%%%%%%%%%%%%%%%%%%%%%%%%%%%%%%%%%
\subsection*{Requirements for Open Data Platforms}

Useful open data platforms should meet a number of essential criteria in order to effectively serve their users. These requirements can be broadly categorized into technical requirements, user-centric requirements, and legal and ethical requirements.

\begin{itemize}[leftmargin=*, label={}]
    \item \textbf{Technical Requirements}:

    \begin{enumerate}
        \item Scalability and Reliability: As a platform's user base expands, it must be able to scale in order to accommodate growing data volumes and user requests. Additionally, it must be trustworthy, with a high availability and robust backup and recovery processes to prevent data loss.
    
        \item Interoperability: The platform must utilize standard data formats and protocols to ensure that the data can be readily accessed and utilized by multiple systems and applications. It should also facilitate the integration and linking of data from various sources.
    
        \item Findability: Effective data indexing and search capabilities are essential for assisting users in locating the required data. This entails offering metadata, unique identifiers, and a robust search engine.
    
        \item Security: The platform must provide robust security measures to prevent unauthorized data access or modification. This includes encryption, access control, and auditing functions.
    \end{enumerate}

    \item \textbf{User-centric Requirements}:

    \begin{enumerate}
        \item Usability: The platform's interface should be simple and intuitive. Users should be able to upload, retrieve, and manipulate data without difficulty.
    
        \item Accessibility: All data should be accessible to users with varying requirements, including those with disabilities. Also, APIs can be provided for developers to access and use the data programmatically.
    
        \item Support: Users should have access to documentation, tutorials, and user support in order to understand how to use the platform and its data.
    \end{enumerate}

    \item \textbf{Legal and Ethical Requirements}:

    \begin{enumerate}
        \item Licensing and Citability: The platform should provide explicit licensing information for each dataset, indicating how it may be utilized and whether or not it should be cited.
    
        \item Compliance with FAIR principles: The data must be discoverable, accessible, interoperable, and reusable (FAIR). This entails making data readily discoverable, accessible under open licenses, compatible with other data sets, and well-documented so that it can be reused in a variety of contexts.
    
        \item Privacy and Ethics: If the platform hosts personal data, it must comply with privacy regulations such as the General Data Protection Regulation (GDPR). Establishing and adhering to ethical guidelines for the collection, use, and sharing of data is necessary.
    
        \item Transparency and Accountability: The platform should have transparent data collection, utilization, and governance policies and be responsible for their enforcement.
    \end{enumerate}

\end{itemize}

%%%%%%%%%%%%%%%%%%%%%%%%%%%%%%%%%%%%%%%%%%%%%%%%%%%%%%%%%%%%%%%%%%%%%%%%%%%%%%%%%%%%%%%%%%%%%%%%%
\subsection*{Challenges and obstacles for Open Data Platforms}

Implementing open data platforms that are widely utilized is a challenging task complicated with numerous challenges and obstacles. While the need for these platforms is widely acknowledged, a number of significant obstacles remain.

\begin{enumerate}
    \item Data Privacy and Security: Ensuring data privacy and security is one of the primary challenges. It is a delicate task to balance the openness of data with the need to safeguard sensitive information. In Europe, laws such as GDPR have imposed stringent data management requirements, making it even more difficult for open data platforms to comply.

    \item Data Standardization and Interoperability: The absence of data standardization is a significant technical obstacle. It can be challenging to integrate disparate data sources onto a single platform due to the diverse formats of the data. This lack of interoperability can reduce the platform's utility, as users may have difficulty locating, comparing, and utilizing data from various sources.

    \item Data integrity and Curation: Another challenge is ensuring the integrity of data on these platforms. Inaccurate data can lead to erroneous conclusions, so open data platforms must have robust data validation and curation procedures. This is a resource-intensive endeavor that frequently necessitates subject-matter specialists and substantial computational resources.

    \item Funding and Sustainability: Open data platforms necessitate substantial resources for development, maintenance, and enhancement. This includes technical infrastructure, personnel, and data curation on an ongoing basis. Obtaining stable, long-term funding for these activities is typically difficult, especially for platforms that provide unrestricted access to their resources.

    \item User Engagement and Training: Lastly, promoting the adoption and proper utilization of open data platforms is a challenge in and of itself. Many potential users may lack the technical expertise required to utilize these platforms effectively. This requires investments in user training and ongoing efforts to enhance the usability and accessibility of the platform.

\end{enumerate}

In conclusion, despite the fact that the implementation of open data platforms can present significant obstacles, the potential benefits to the scientific community and the general public make it worthwhile to overcome these obstacles. To overcome these obstacles and actualize the full potential of open data, a coordinated effort involving policymakers, funding bodies, technical experts, and end users is required.

%%%%%%%%%%%%%%%%%%%%%%%%%%%%%%%%%%%%%%%%%%%%%%%%%%%%%%%%%%%%%%%%%%%%%%%%%%%%%%%%%%%%%%%%%%%%%%%%%
\subsection*{Lessons Learned}

%%% ADD MAYBE A SUMMARY AS IN: If you are a normal math person, then you need repo A. If you additionally want this and that then you need Repo B. ...

The sharing of research data is essential for scientific research in all domains and disciplines. Open data platforms promote transparency, reproducibility, and the acceleration of scientific discovery by adhering to Open Science principles. In mathematics, open data platforms are becoming increasingly important for researchers, as they facilitate the sharing and reuse of research data. In this article, we gave an overview and an evaluation of the current status of open data platforms, with a focus on the field of mathematics. We included their primary requirements and the obstacles preventing their successful implementation.

To be successful, an open data platform must satisfy numerous requirements that we addressed in this article. However, also several challenges and obstacles were identified, including:

\begin{enumerate}

    \item Ensure that all researchers, regardless of their location, resources, or expertise, have easy access to research data in order to promote data sharing. However, some open data platforms may be difficult to access, posing obstacles for researchers who wish to use the platform.

    \item For encouraging researchers to share and utilize data, user-friendly platforms are essential. Researchers may be less inclined to use a platform if it is too complicated or difficult to navigate.

    \item For researchers who want to share their data on an open platform, efficient data submission protocols are essential. Processes that are cumbersome or time-consuming may discourage researchers from submitting their data, thereby diminishing the efficacy of the platform.

    \item Open data platforms must adhere to the FAIR principles to guarantee that the data is discoverable, accessible, interoperable, and reusable. Platforms that do not adhere to these recommendations may reduce the utility of the data shared on their platform.
    
\end{enumerate}

Understanding these challenges and obstacles can help researchers and developers to create - or improve - more effective platforms for sharing mathematical research data. This way, open data platforms can become an even more valuable resource for mathematicians, fostering collaboration, the sharing of data, and the acceleration of scientific discovery.

%%%%%%%%%%%%%%%%%%%%%%%%%%%%%%%%%%%%%%%%%%%%%%%%%%%%%%%%%%%%%%%%%%%%%%%%%%%%%%%%%%%%%%%%%%%%%%%%%
\section*{APPENDIX A: FAIR principles}
\label{sec:fair-principles}
%%%%%%%%%%%%%%%%%%%%%%%%%%%%%%%%%%%%%%%%%%%%%%%%%%%%%%%%%%%%%%%%%%%%%%%%%%%%%%%%%%%%%%%%%%%%%%%%%

The adherence to the FAIR principles is one of the criteria that we have used to evaluate the current status of open data platforms in the field of mathematics. These guiding principles prioritize machine-actionability and emphasize the findability, accessibility, interoperability, and reusability of data. The primary objective of these principles is to facilitate the sustainable reuse of research data by establishing guidelines that research data platforms and infrastructures should adhere to as part of their service offerings.

These principles were initially introduced in a publication by Wilkinson et al. in 2016~\cite{wilkinson2016fair}. For a thorough discussion on their interpretation and additional implementation considerations, refer to the work of Jacobsen et al.~\cite{jacobsen_fair} For completeness, we present below the list of FAIR principles as outlined by Wilkinson et al. ~\cite{wilkinson2016fair}

\begin{itemize}

    \item Findability
    \begin{itemize}
        \item F1: (Meta)data are assigned globally unique and persistent identifiers.
        \item F2: Data are described with rich metadata.    
        \item F3: Metadata clearly and explicitly include the identifier of the data it describes.
        \item F4: (Meta)data are registered or indexed in a searchable resource.
    \end{itemize}

    \item Accessibility
    \begin{itemize}
        \item A1: (Meta)data are retrievable by their identifier using a standardized communication protocol.
        \item A1.1: The protocol is open, free and universally implementable.
        \item A1.2: The protocol allows for an authentication and authorisation procedure, where necessary.
        \item A2: Metadata are accessible, even when the data are no longer available
    \end{itemize}

    \item Interoperability
    \begin{itemize}
        \item I1: (Meta)data use a formal, accessible, shared, and broadly applicable language for knowledge representation.
        \item I2: (Meta)data use vocabularies that follow the FAIR principles.
        \item I3: (Meta)data include qualified references to other (meta)data.
    \end{itemize}

    \item Reusability
    \begin{itemize}
        \item R1: (Meta)data are richly described with a plurality of accurate and relevant attributes.
        \item R1.1: (Meta)data are released with a clear and accessible data usage license.
        \item R1.2: (Meta)data are associated with detailed provenance.
        \item R1.3: (Meta)data meet domain-relevant community standards.
    \end{itemize}

\end{itemize}

In summary, the findability principles emphasize the indexing of metadata using persistent and unique identifiers, enabling the discovery and unique identification of data resources. The accessibility principles refer to the capability to access data via open protocols after it has been located, which may involve authentication and authorization. The interoperability principles specifically address machine-actionability, so that data can be readily integrated into existing workflows. The reusability principles describe the criteria for enriching metadata, enabling users to identify the specific context and conditions under which the data can be re-used.

%%%%%%%%%%%%%%%%%%%%%%%%%%%%%%%%%%%%%%%%%%%%%%%%%%%%%%%%%%%%%%%%%%%%%%%%%%%%%%%%%%%%%%%%%%%%%%%%%
\section*{APPENDIX B: MediaWiki Math Platforms}
\label{sec:mediawiki-math}
%%%%%%%%%%%%%%%%%%%%%%%%%%%%%%%%%%%%%%%%%%%%%%%%%%%%%%%%%%%%%%%%%%%%%%%%%%%%%%%%%%%%%%%%%%%%%%%%%

Mathematical research data is frequently published online through wiki-based platforms. These platforms allow for the rapid creation and iterative editing of entries in a collaborative environment. Due to the convenience of this approach, there are several platforms dedicated to specific mathematical disciplines. Table~\ref{tab:mediawiki_platforms} in this appendix presents a selection of exemplary platforms that fall into this category.

The table includes a collection of platforms, all of which operate on MediaWiki, except for nLab, which is built on Instiki, a wiki software based on Ruby on Rails. Among the platforms listed, there are a few noteworthy examples worth highlighting: the Encyclopedia of Math, Complexity Zoo, and nLab. The Encyclopedia of Math is an online wiki initially created by the Springer Verlag and managed in cooperation with the European Mathematical Society. It hosts over 8,000 articles covering advanced mathematical topics, which can be updated by users and undergo editorial board review for accuracy. The Complexity Zoo, initiated by Scott Aaronson in 2002, aims to catalog all classes of computational complexity and currently documents over 500 complexity classes. Lastly, nLab is a collaborative platform that includes more than 18,000 pages spanning mathematics, physics, and philosophy, with a strong emphasis on type theory, category theory, and homotopy theory.

\begin{table}[ht]
\centering
\newcolumntype{R}{>{\raggedright\arraybackslash}p{5cm}}
\begin{tabularx}{16cm}{p{4cm} p{5.5cm} R }
\toprule
Platforms & Math Focus & URL \\
\midrule
Boolean Zoo          & Boolean analysis                      & \href{https://booleanzoo.weizmann.ac.il/}{booleanzoo.weizmann.ac.il} \\ \hline
Complexity Zoo       & Complexity classes                    & \href{https://complexityzoo.net/}{complexityzoo.net} \\ \hline
Encyclopedia of Math & Articles in the field of mathematics  & \href{https://encyclopediaofmath.org/}{encyclopediaofmath.org} \\ \hline
GroupProps           & Group properties                      & \href{https://groupprops.subwiki.org}{groupprops.subwiki.org} \\ \hline
MOR Wiki             & Model benchmarks                      & \href{https://morwiki.mpi-magdeburg.mpg.de/}{morwiki.mpi-magdeburg.mpg.de} \\ \hline
nLab                 & Category theory                       & \href{https://ncatlab.org/}{ncatlab.org} \\ \hline
The Knot Atlas       & Knots                                 & \href{http://katlas.org/}{katlas.org} \\ \hline
The Manifold Atlas   & Manifolds                             & \href{http://http://www.map.mpim-bonn.mpg.de/}{map.mpim-bonn.mpg.de} \\ 
\bottomrule
\end{tabularx}
\caption{\label{tab:mediawiki_platforms}Mathematical research data platforms based on MediaWiki}
\end{table}

While this approach facilitates the sharing of research data, the existing implementations currently show limited adherence to the FAIR principles. Merely relying on the default environment offered by a MediaWiki instance does not inherently ensure satisfactory compliance with FAIR principles. Achieving proper FAIR adherence requires a deliberate effort and the addition of supplementary measures.

Most of the platforms featured in this table do not assign persistent identifiers to their resources. Instead, the resources are solely accessible through URLs, without any guarantee of persistence. Despite the availability of MediaWiki's API for accessing metadata on existing MediaWiki sites, the listed platforms do not use this mechanism to provide comprehensive metadata describing their stored resources. Qualified references to other resources are rarely provided, and explicit license information is seldom included. Only in cases where user pages are available for each user, some individuals may voluntarily add identifiers, such as an ORCID ID, which can serve as qualified references. However, this information is not directly included in the retrieved metadata.

In summary, using MediaWiki as the foundation for a research data platform can serve as a valuable initial step in establishing a collaborative environment for resource sharing among users. However, to ensure adherence to the FAIR principles, additional features must be implemented beyond the basic configuration. These include assigning persistent identifiers to resources, making these identifiers readily available along with comprehensive and contextual metadata through the provided API, implementing controlled vocabularies, including qualified references to other resources, and clearly publishing license information.

\bibliography{biblio}

\begin{thebibliography}{10}
\urlstyle{rm}
\expandafter\ifx\csname url\endcsname\relax
  \def\url#1{\texttt{#1}}\fi
\expandafter\ifx\csname urlprefix\endcsname\relax\def\urlprefix{URL }\fi
\expandafter\ifx\csname doiprefix\endcsname\relax\def\doiprefix{DOI: }\fi
\providecommand{\bibinfo}[2]{#2}
\providecommand{\eprint}[2][]{\url{#2}}

\bibitem{Piwowar2007}
\bibinfo{author}{Piwowar, H.~A.}, \bibinfo{author}{Day, R.~S.} \& \bibinfo{author}{Fridsma, D.~B.}
\newblock \bibinfo{journal}{\bibinfo{title}{Sharing detailed research data is associated with increased citation rate}}.
\newblock {\emph{\JournalTitle{PLOS ONE}}} \textbf{\bibinfo{volume}{2}}, \bibinfo{pages}{1--5}, \url{10.1371/journal.pone.0000308} (\bibinfo{year}{2007}).

\bibitem{Tenopir2015}
\bibinfo{author}{Tenopir, C.} \emph{et~al.}
\newblock \bibinfo{journal}{\bibinfo{title}{Changes in data sharing and data reuse practices and perceptions among scientists worldwide}}.
\newblock {\emph{\JournalTitle{PLOS ONE}}} \textbf{\bibinfo{volume}{10}}, \bibinfo{pages}{1--24}, \url{10.1371/journal.pone.0134826} (\bibinfo{year}{2015}).

\bibitem{Lebo2018}
\bibinfo{author}{Lebo, M.~S.} \emph{et~al.}
\newblock \bibinfo{journal}{\bibinfo{title}{Data sharing as a national quality improvement program: reporting on brca1 and brca2 variant-interpretation comparisons through the canadian open genetics repository (cogr)}}.
\newblock {\emph{\JournalTitle{Genetics in Medicine}}} \textbf{\bibinfo{volume}{20}}, \bibinfo{pages}{294--302}, \url{10.1038/gim.2017.80} (\bibinfo{year}{2018}).

\bibitem{Thoegersen2022}
\bibinfo{author}{Thoegersen, J.} \& \bibinfo{author}{Borlund, P.}
\newblock \bibinfo{journal}{\bibinfo{title}{Researcher attitudes toward data sharing in public data repositories: a meta-evaluation of studies on researcher data sharing}}.
\newblock {\emph{\JournalTitle{Journal of Documentation}}} \textbf{\bibinfo{volume}{78}}, \bibinfo{pages}{1--17}, \url{10.1108/JD-01-2021-0015} (\bibinfo{year}{2022}).

\bibitem{pampel2013making}
\bibinfo{author}{Pampel, H.} \emph{et~al.}
\newblock \bibinfo{journal}{\bibinfo{title}{Making research data repositories visible: the re3data. org registry}}.
\newblock {\emph{\JournalTitle{PloS one}}} \textbf{\bibinfo{volume}{8}}, \bibinfo{pages}{e78080} (\bibinfo{year}{2013}).

\bibitem{DBLP:journals/pami/GreinerPetterSBSAG23}
\bibinfo{author}{Greiner{-}Petter, A.} \emph{et~al.}
\newblock \bibinfo{journal}{\bibinfo{title}{Do the math: Making mathematics in wikipedia computable}}.
\newblock {\emph{\JournalTitle{{IEEE} Trans. Pattern Anal. Mach. Intell.}}} \textbf{\bibinfo{volume}{45}}, \bibinfo{pages}{4384--4395}, \url{10.1109/TPAMI.2022.3195261} (\bibinfo{year}{2023}).

\bibitem{DBLP:conf/mkm/CohlGS18}
\bibinfo{author}{Cohl, H.~S.}, \bibinfo{author}{Greiner{-}Petter, A.} \& \bibinfo{author}{Schubotz, M.}
\newblock \bibinfo{title}{Automated symbolic and numerical testing of {DLMF} formulae using computer algebra systems}.
\newblock In \bibinfo{editor}{Rabe, F.}, \bibinfo{editor}{Farmer, W.~M.}, \bibinfo{editor}{Passmore, G.~O.} \& \bibinfo{editor}{Youssef, A.} (eds.) \emph{\bibinfo{booktitle}{Intelligent Computer Mathematics - 11th International Conference, {CICM} 2018, Hagenberg, Austria, August 13-17, 2018, Proceedings}}, vol. \bibinfo{volume}{11006} of \emph{\bibinfo{series}{Lecture Notes in Computer Science}}, \bibinfo{pages}{39--52}, \url{10.1007/978-3-319-96812-4\_4} (\bibinfo{publisher}{Springer}, \bibinfo{year}{2018}).

\bibitem{boulton2012}
\bibinfo{author}{Boulton, G.}
\newblock \bibinfo{journal}{\bibinfo{title}{Open your minds and share your results}}.
\newblock {\emph{\JournalTitle{Nature}}} \textbf{\bibinfo{volume}{486}} (\bibinfo{year}{2012}).

\bibitem{mckiernan2016open}
\bibinfo{author}{McKiernan, E.~C.} \emph{et~al.}
\newblock \bibinfo{journal}{\bibinfo{title}{How open science helps researchers succeed}}.
\newblock {\emph{\JournalTitle{elife}}} \textbf{\bibinfo{volume}{5}}, \bibinfo{pages}{e16800} (\bibinfo{year}{2016}).

\bibitem{DBLP:phd/dnb/Schubotz17}
\bibinfo{author}{Schubotz, M.}
\newblock \emph{\bibinfo{title}{Augmenting Mathematical Formulae for More Effective Querying {\&} Efficient Presentation}}.
\newblock Ph.D. thesis, \bibinfo{school}{Technical University of Berlin, Germany} (\bibinfo{year}{2017}).

\bibitem{sansone2019fairsharing}
\bibinfo{author}{Sansone, S.-A.} \emph{et~al.}
\newblock \bibinfo{journal}{\bibinfo{title}{Fairsharing as a community approach to standards, repositories and policies}}.
\newblock {\emph{\JournalTitle{Nature biotechnology}}} \textbf{\bibinfo{volume}{37}}, \bibinfo{pages}{358--367} (\bibinfo{year}{2019}).

\bibitem{iancu2014system}
\bibinfo{author}{Iancu, M.}, \bibinfo{author}{Jucovschi, C.}, \bibinfo{author}{Kohlhase, M.} \& \bibinfo{author}{Wiesing, T.}
\newblock \bibinfo{title}{System description: Mathhub. info}.
\newblock In \emph{\bibinfo{booktitle}{International Conference on Intelligent Computer Mathematics}}, \bibinfo{pages}{431--434} (\bibinfo{organization}{Springer}, \bibinfo{year}{2014}).

\bibitem{Re3data}
\bibinfo{author}{Pampel, H.} \emph{et~al.}
\newblock \bibinfo{journal}{\bibinfo{title}{Making {Research} {Data} {Repositories} {Visible}: The re3data.org {Registry}}}.
\newblock {\emph{\JournalTitle{PLOS ONE}}} \textbf{\bibinfo{volume}{8}}, \bibinfo{pages}{e78080} (\bibinfo{year}{2013}).

\bibitem{narboux2016towards}
\bibinfo{author}{Narboux, J.} \& \bibinfo{author}{Braun, D.}
\newblock \bibinfo{journal}{\bibinfo{title}{Towards a certified version of the encyclopedia of triangle centers}}.
\newblock {\emph{\JournalTitle{Mathematics in Computer Science}}} \textbf{\bibinfo{volume}{10}}, \bibinfo{pages}{57--73} (\bibinfo{year}{2016}).

\bibitem{de2016information}
\bibinfo{author}{de~Ridder, H.~N.} \emph{et~al.}
\newblock \bibinfo{title}{Information system on graph classes and their inclusions (isgci)}.
\newblock \bibinfo{howpublished}{\url{https://www.graphclasses.org/}}.
\newblock \bibinfo{note}{Accessed: 2023-06-13}.

\bibitem{brown2015graded}
\bibinfo{author}{Brown, G.}, \bibinfo{author}{Kasprzyk, A.~M.} \emph{et~al.}
\newblock \bibinfo{title}{Graded ring database}.
\newblock \bibinfo{howpublished}{\url{http://www.grdb.co.uk/}}.
\newblock \bibinfo{note}{Accessed: 2023-06-13}.

\bibitem{cremona2016functions}
\bibinfo{author}{Cremona, J.}
\newblock \bibinfo{journal}{\bibinfo{title}{The l-functions and modular forms database project}}.
\newblock {\emph{\JournalTitle{Foundations of Computational Mathematics}}} \textbf{\bibinfo{volume}{16}}, \bibinfo{pages}{1541--1553} (\bibinfo{year}{2016}).

\bibitem{neumann2014datacite}
\bibinfo{author}{Neumann, J.} \& \bibinfo{author}{Brase, J.}
\newblock \bibinfo{journal}{\bibinfo{title}{Datacite and doi names for research data}}.
\newblock {\emph{\JournalTitle{Journal of computer-aided molecular design}}} \textbf{\bibinfo{volume}{28}}, \bibinfo{pages}{1035--1041} (\bibinfo{year}{2014}).

\bibitem{bode2018guide}
\bibinfo{author}{Bode, C.}, \bibinfo{author}{Herzog, C.}, \bibinfo{author}{Hook, D.} \& \bibinfo{author}{McGrath, R.}
\newblock \bibinfo{journal}{\bibinfo{title}{A guide to the dimensions data approach}}.
\newblock {\emph{\JournalTitle{Dimensions Report. Cambridge, MA: Digital Science}}}  (\bibinfo{year}{2018}).

\bibitem{ardestani2015b2share}
\bibinfo{author}{Ardestani, S.~B.} \emph{et~al.}
\newblock \bibinfo{title}{B2share: An open escience data sharing platform}.
\newblock In \emph{\bibinfo{booktitle}{2015 IEEE 11th International Conference on e-Science}}, \bibinfo{pages}{448--453} (\bibinfo{organization}{IEEE}, \bibinfo{year}{2015}).

\bibitem{white2008dryad}
\bibinfo{author}{White, H.}, \bibinfo{author}{Carrier, S.}, \bibinfo{author}{Thompson, A.}, \bibinfo{author}{Greenberg, J.} \& \bibinfo{author}{Scherle, R.}
\newblock \bibinfo{title}{The dryad data repository: A singapore framework metadata architecture in a dspace environment.}
\newblock In \emph{\bibinfo{booktitle}{Dublin core conference}}, \bibinfo{pages}{157--162} (\bibinfo{year}{2008}).

\bibitem{wolstencroft2017fairdomhub}
\bibinfo{author}{Wolstencroft, K.} \emph{et~al.}
\newblock \bibinfo{journal}{\bibinfo{title}{Fairdomhub: a repository and collaboration environment for sharing systems biology research}}.
\newblock {\emph{\JournalTitle{Nucleic acids research}}} \textbf{\bibinfo{volume}{45}}, \bibinfo{pages}{D404--D407} (\bibinfo{year}{2017}).

\bibitem{bhoi2018mendeley}
\bibinfo{author}{Bhoi, N.~K.}
\newblock \bibinfo{journal}{\bibinfo{title}{Mendeley data repository as a platform for research data management}}.
\newblock {\emph{\JournalTitle{Marching beyond libraries: Managerial skills and technological competencies}}} \bibinfo{pages}{481--487} (\bibinfo{year}{2018}).

\bibitem{li2019moving}
\bibinfo{author}{Li, R.} \emph{et~al.}
\newblock \bibinfo{journal}{\bibinfo{title}{Moving data sharing forward: the launch of the vivli platform}}.
\newblock {\emph{\JournalTitle{NAM Perspectives}}} \textbf{\bibinfo{volume}{8}} (\bibinfo{year}{2019}).

\bibitem{Austin2016Research}
\bibinfo{author}{Austin, C.~C.} \emph{et~al.}
\newblock \bibinfo{journal}{\bibinfo{title}{Research {Data} {Repositories}: Review of {Current} {Features}, {Gap} {Analysis}, and {Recommendations} for {Minimum} {Requirements}}}.
\newblock {\emph{\JournalTitle{IASSIST quarterly}}} \textbf{\bibinfo{volume}{39}}, \bibinfo{pages}{24}, \url{10.29173/IQ904} (\bibinfo{year}{2015}).

\bibitem{wilkinson2016fair}
\bibinfo{author}{Wilkinson, M.~D.} \emph{et~al.}
\newblock \bibinfo{journal}{\bibinfo{title}{The fair guiding principles for scientific data management and stewardship}}.
\newblock {\emph{\JournalTitle{Scientific data}}} \textbf{\bibinfo{volume}{3}}, \bibinfo{pages}{1--9} (\bibinfo{year}{2016}).

\bibitem{cruz2018adding}
\bibinfo{author}{Cruz, M.~J.} \& \bibinfo{author}{Gramsbergen, E.}
\newblock \bibinfo{title}{Adding value and facilitating data reuse: the case of the 4tu. centre for research data}.
\newblock In \emph{\bibinfo{booktitle}{Proceedings of the PV2018 Conference, 15-17 May 2018, Harwell, UK}} (\bibinfo{year}{2018}).

\bibitem{Blanchette2015Mining}
\bibinfo{author}{Blanchette, J.~C.}, \bibinfo{author}{Haslbeck, M.}, \bibinfo{author}{Matichuk, D.} \& \bibinfo{author}{Nipkow, T.}
\newblock \bibinfo{title}{{Mining the Archive of Formal Proofs}}.
\newblock In \emph{\bibinfo{booktitle}{{CICM 2015}}}, Intelligent Computer Mathematics - International Conference, CICM 2015, Washington, DC, USA, July 13-17, 2015, Proceedings, \url{10.1007/978-3-319-20615-8\_1} (\bibinfo{address}{Washington DC, United States}, \bibinfo{year}{2015}).

\bibitem{MacKenzie2022Re}
\bibinfo{author}{MacKenzie, C.}, \bibinfo{author}{Huch, F.}, \bibinfo{author}{Vaughan, J.} \& \bibinfo{author}{Fleuriot, J.}
\newblock \emph{\bibinfo{title}{Re-imagining the {Isabelle} {Archive} of {Formal} {Proofs}}}, \bibinfo{pages}{162--167} (\bibinfo{year}{2022}).

\bibitem{Paulson1989}
\bibinfo{author}{Paulson, L.}
\newblock \bibinfo{journal}{\bibinfo{title}{The foundation of a generic theorem prover}}.
\newblock {\emph{\JournalTitle{J Autom Reasoning}}} \textbf{\bibinfo{volume}{5}} (\bibinfo{year}{1989}).

\bibitem{malik2020biomodels}
\bibinfo{author}{Malik-Sheriff, R.~S.} \emph{et~al.}
\newblock \bibinfo{journal}{\bibinfo{title}{Biomodels—15 years of sharing computational models in life science}}.
\newblock {\emph{\JournalTitle{Nucleic acids research}}} \textbf{\bibinfo{volume}{48}}, \bibinfo{pages}{D407--D415} (\bibinfo{year}{2020}).

\bibitem{Schwiebert_RingApp_2023}
\bibinfo{author}{Schwiebert, R.}
\newblock \bibinfo{title}{Ringapp v1.1.0}.
\newblock \bibinfo{howpublished}{\url{https://github.com/rschwiebert/RingApp}}.
\newblock \bibinfo{note}{Accessed: 2023-06-13}.

\bibitem{Thelwall2016figshare}
\bibinfo{author}{Thelwall, M.} \& \bibinfo{author}{Kousha, K.}
\newblock \bibinfo{journal}{\bibinfo{title}{Figshare: a universal repository for academic resource sharing?}}
\newblock {\emph{\JournalTitle{Online Information Review}}} \textbf{\bibinfo{volume}{40}}, \bibinfo{pages}{333--346}, \url{10.1108/OIR-06-2015-0190} (\bibinfo{year}{2016}).

\bibitem{berg2014findstat}
\bibinfo{author}{Berg, C.}, \bibinfo{author}{Pons, V.}, \bibinfo{author}{Scrimshaw, T.}, \bibinfo{author}{Striker, J.} \& \bibinfo{author}{Stump, C.}
\newblock \bibinfo{title}{Findstat - the combinatorial statistics database} (\bibinfo{year}{2014}).
\newblock \eprint{1401.3690}.

\bibitem{Baruch2007open}
\bibinfo{author}{Baruch, P.}
\newblock \bibinfo{journal}{\bibinfo{title}{Open access developments in france: the {HAL} open archives system}}.
\newblock {\emph{\JournalTitle{Learned Publishing}}} \textbf{\bibinfo{volume}{20}}, \bibinfo{pages}{267--282}, \url{10.1087/095315107X239636} (\bibinfo{year}{2007}).

\bibitem{magazine2011dataverse}
\bibinfo{author}{Magazine, D.-L.}
\newblock \bibinfo{journal}{\bibinfo{title}{The dataverse network{\textregistered}: an open-source application for sharing, discovering and preserving data}}.
\newblock {\emph{\JournalTitle{D-lib Magazine}}} \textbf{\bibinfo{volume}{17}} (\bibinfo{year}{2011}).

\bibitem{fevola2022mathematical}
\bibinfo{author}{Fevola, C.} \& \bibinfo{author}{Görgen, C.}
\newblock \bibinfo{title}{The mathematical research-data repository mathrepo} (\bibinfo{year}{2022}).
\newblock \eprint{2202.04022}.

\bibitem{Rossi2015network}
\bibinfo{author}{Rossi, R.} \& \bibinfo{author}{Ahmed, N.}
\newblock \bibinfo{title}{The network data repository with interactive graph analytics and visualization}.
\newblock In \emph{\bibinfo{booktitle}{Proceedings of the AAAI conference on artificial intelligence}}, vol.~\bibinfo{volume}{29} (\bibinfo{year}{2015}).

\bibitem{sloane2003line}
\bibinfo{author}{Sloane, N.~J.} \emph{et~al.}
\newblock \bibinfo{title}{The on-line encyclopedia of integer sequences} (\bibinfo{year}{2003}).

\bibitem{foster2017open}
\bibinfo{author}{Foster, E.~D.} \& \bibinfo{author}{Deardorff, A.}
\newblock \bibinfo{journal}{\bibinfo{title}{Open science framework (osf)}}.
\newblock {\emph{\JournalTitle{Journal of the Medical Library Association: JMLA}}} \textbf{\bibinfo{volume}{105}}, \bibinfo{pages}{203}, \url{10.5195/jmla.2017.88} (\bibinfo{year}{2017}).

\bibitem{pi-base-data}
\bibinfo{title}{{pi-Base Community, pi-Base/data}}.
\newblock \bibinfo{howpublished}{\url{https://github.com/pi-base}}.
\newblock \bibinfo{note}{Accessed: 2023-06-13}.

\bibitem{paffenholz2017polydb}
\bibinfo{author}{Paffenholz, A.}
\newblock \bibinfo{title}{polydb: A database for polytopes and related objects} (\bibinfo{year}{2017}).
\newblock \eprint{1711.02936}.

\bibitem{Chenzan2017sciencedb}
\bibinfo{author}{Chengzan, L.}, \bibinfo{author}{Yanfei, H.}, \bibinfo{author}{Jianhui, L.} \& \bibinfo{author}{Lili, Z.}
\newblock \bibinfo{title}{{ScienceDB}: A public multidisciplinary research data repository for {eScience}}.
\newblock In \emph{\bibinfo{booktitle}{2017 {IEEE} 13th International Conference on e-Science (e-Science)}}, \bibinfo{pages}{248--255}, \url{10.1109/eScience.2017.38} (\bibinfo{year}{2017}).

\bibitem{kolodziej2019suitesparse}
\bibinfo{author}{Kolodziej, S.~P.} \emph{et~al.}
\newblock \bibinfo{journal}{\bibinfo{title}{The suitesparse matrix collection website interface}}.
\newblock {\emph{\JournalTitle{Journal of Open Source Software}}} \textbf{\bibinfo{volume}{4}}, \bibinfo{pages}{1244} (\bibinfo{year}{2019}).

\bibitem{coolsaet23}
\bibinfo{author}{Coolsaet, K.}, \bibinfo{author}{D’hondt, S.} \& \bibinfo{author}{Goedgebeur, J.}
\newblock \bibinfo{journal}{\bibinfo{title}{House of graphs 2.0: A database of interesting graphs and more}}.
\newblock {\emph{\JournalTitle{Discrete Applied Mathematics}}} \textbf{\bibinfo{volume}{325}}, \bibinfo{pages}{97--107}, \url{10.1016/j.dam.2022.10.013} (\bibinfo{year}{2023}).

\bibitem{Vrandecic2014Wikidata}
\bibinfo{author}{Vrande{\v c}i{\' c}, D.} \& \bibinfo{author}{Kr{\" o}tzsch, M.}
\newblock \bibinfo{journal}{\bibinfo{title}{Wikidata}}.
\newblock {\emph{\JournalTitle{Communications of the ACM}}} \textbf{\bibinfo{volume}{57}}, \bibinfo{pages}{78--85} (\bibinfo{year}{2014}).

\bibitem{DBLP:conf/mkm/SchubotzW14}
\bibinfo{author}{Schubotz, M.} \& \bibinfo{author}{Wicke, G.}
\newblock \bibinfo{title}{Mathoid: Robust, scalable, fast and accessible math rendering for wikipedia}.
\newblock In \bibinfo{editor}{Watt, S.~M.}, \bibinfo{editor}{Davenport, J.~H.}, \bibinfo{editor}{Sexton, A.~P.}, \bibinfo{editor}{Sojka, P.} \& \bibinfo{editor}{Urban, J.} (eds.) \emph{\bibinfo{booktitle}{Intelligent Computer Mathematics - International Conference, {CICM} 2014, Coimbra, Portugal, July 7-11, 2014. Proceedings}}, vol. \bibinfo{volume}{8543} of \emph{\bibinfo{series}{Lecture Notes in Computer Science}}, \bibinfo{pages}{224--235}, \url{10.1007/978-3-319-08434-3\_17} (\bibinfo{publisher}{Springer}, \bibinfo{year}{2014}).

\bibitem{DBLP:conf/jcdl/SchubotzGMTG20}
\bibinfo{author}{Schubotz, M.}, \bibinfo{author}{Greiner{-}Petter, A.}, \bibinfo{author}{Meuschke, N.}, \bibinfo{author}{Teschke, O.} \& \bibinfo{author}{Gipp, B.}
\newblock \bibinfo{title}{Mathematical formulae in wikimedia projects 2020}.
\newblock In \bibinfo{editor}{Huang, R.} \emph{et~al.} (eds.) \emph{\bibinfo{booktitle}{{JCDL} '20: Proceedings of the {ACM/IEEE} Joint Conference on Digital Libraries in 2020, Virtual Event, China, August 1-5, 2020}}, \bibinfo{pages}{447--448}, \url{10.1145/3383583.3398557} (\bibinfo{publisher}{{ACM}}, \bibinfo{year}{2020}).

\bibitem{Waagmeester2020Wikidata}
\bibinfo{author}{Waagmeester, A.} \emph{et~al.}
\newblock \bibinfo{journal}{\bibinfo{title}{Wikidata as a knowledge graph for the life sciences}}.
\newblock {\emph{\JournalTitle{eLife}}} \textbf{\bibinfo{volume}{9}} (\bibinfo{year}{2020}).

\bibitem{DBLP:conf/semweb/ScharpfSG21}
\bibinfo{author}{Scharpf, P.}, \bibinfo{author}{Schubotz, M.} \& \bibinfo{author}{Gipp, B.}
\newblock \bibinfo{title}{Mathematics in wikidata}.
\newblock In \bibinfo{editor}{Kaffee, L.}, \bibinfo{editor}{Razniewski, S.} \& \bibinfo{editor}{Hogan, A.} (eds.) \emph{\bibinfo{booktitle}{Proceedings of the 2nd Wikidata Workshop (Wikidata 2021) co-located with the 20th International Semantic Web Conference {(ISWC} 2021), Virtual Conference, October 24, 2021}}, vol. \bibinfo{volume}{2982} of \emph{\bibinfo{series}{{CEUR} Workshop Proceedings}} (\bibinfo{publisher}{CEUR-WS.org}, \bibinfo{year}{2021}).

\bibitem{DBLP:conf/sigir/ScharpfSG18}
\bibinfo{author}{Scharpf, P.}, \bibinfo{author}{Schubotz, M.} \& \bibinfo{author}{Gipp, B.}
\newblock \bibinfo{title}{Representing mathematical formulae in content mathml using wikidata}.
\newblock In \bibinfo{editor}{Mayr, P.}, \bibinfo{editor}{Chandrasekaran, M.~K.} \& \bibinfo{editor}{Jaidka, K.} (eds.) \emph{\bibinfo{booktitle}{Proceedings of the 3rd Joint Workshop on Bibliometric-enhanced Information Retrieval and Natural Language Processing for Digital Libraries {(BIRNDL} 2018) co-located with the 41st International {ACM} {SIGIR} Conference on Research and Development in Information Retrieval {(SIGIR} 2018), Ann Arbor, USA, July 12, 2018}}, vol. \bibinfo{volume}{2132} of \emph{\bibinfo{series}{{CEUR} Workshop Proceedings}}, \bibinfo{pages}{46--59} (\bibinfo{publisher}{CEUR-WS.org}, \bibinfo{year}{2018}).

\bibitem{https://doi.org/10.25495/7gxk-rd71}
\bibinfo{author}{{European Organization For Nuclear Research}} \& \bibinfo{author}{{OpenAIRE}}.
\newblock \bibinfo{title}{Zenodo}, \url{10.25495/7GXK-RD71} (\bibinfo{year}{2013}).

\bibitem{jacobsen_fair}
\bibinfo{author}{Jacobsen, A.} \emph{et~al.}
\newblock \bibinfo{journal}{\bibinfo{title}{{FAIR Principles: Interpretations and Implementation Considerations}}}.
\newblock {\emph{\JournalTitle{Data Intelligence}}} \textbf{\bibinfo{volume}{2}}, \bibinfo{pages}{10--29}, \url{10.1162/dint_r_00024} (\bibinfo{year}{2020}).
\newblock \eprint{https://direct.mit.edu/dint/article-pdf/2/1-2/10/1893430/dint\_r\_00024.pdf}.

\end{thebibliography}

\section*{Acknowledgements} 

We would like to thank Dr. Olaf Teschke for helpful comments and suggestions for this manuscript. This work was supported by the Deutsche Forschungsgemeinschaft (DFG) (project grant 460135501).

\section*{Author contributions statement}

TC wrote the first draft of the manuscript. All authors contributed equally to the literature review and to writing the manuscript. All authors reviewed the manuscript. 

\section*{Competing interests}

The authors declare no competing interests.

\section*{Data Availability}

No datasets were generated during this study.

\section*{Code Availability}

No codes were generated during this study.

\end{document}